\documentclass[%
 reprint,
superscriptaddress,
 amsmath,amssymb,
 aps,
prb,
]{revtex4-2}

\usepackage{graphicx}
\usepackage{dcolumn}
\usepackage{bm}
\usepackage{svg}
\usepackage{microtype} 
\usepackage{paralist}
\usepackage[normalem]{ulem}
\usepackage[unicode=true, colorlinks=true, citecolor={blue!80!black}, urlcolor={blue!50!black}, linkcolor = {blue!80!black}]{hyperref} 
\usepackage{array}
\usepackage{siunitx}
\usepackage{wrapfig}
\usepackage{enumitem}
\usepackage{amsmath,amssymb}
\usepackage{array}
\usepackage{soul}
\usepackage{centernot}
\usepackage{bibunits}
\usepackage{romannum}
\usepackage[titletoc, title]{appendix}
\usepackage{titletoc}
\usepackage{lineno}
\usepackage{lipsum}  
\usepackage{printlen}

\newcommand*{\balancecolsandclearpage}{%
  \cleardoublepage
  \twocolumngrid
}

\newcommand{\alox}{AlO\textsubscript{x}}

\begin{document}
\pagenumbering{arabic}

\preprint{APS/123-QED}

\title{Andreev spin relaxation time in a shadow-evaporated InAs weak link}

\author{Haoran Lu}
\email{hl2396@cornell.edu}
\affiliation{School of Applied and Engineering Physics, Cornell University, Ithaca, NY, 14853, USA}

\author{David F. Bofill}
\affiliation{Center for Quantum Devices, Niels Bohr Institute, University of Copenhagen, 2100 Copenhagen, Denmark}
\affiliation{NNF Quantum Computing Programme, Niels Bohr Institute, University of Copenhagen, 2100 Copenhagen, Denmark}

\author{Zhenhai Sun}
\affiliation{Center for Quantum Devices, Niels Bohr Institute, University of Copenhagen, 2100 Copenhagen, Denmark}
\affiliation{NNF Quantum Computing Programme, Niels Bohr Institute, University of Copenhagen, 2100 Copenhagen, Denmark}

\author{Thomas Kanne }
\affiliation{Center for Quantum Devices, Niels Bohr Institute, University of Copenhagen, 2100 Copenhagen, Denmark}

\author{Jesper Nygård}
\affiliation{Center for Quantum Devices, Niels Bohr Institute, University of Copenhagen, 2100 Copenhagen, Denmark}

\author{Morten Kjaergaard}
\affiliation{Center for Quantum Devices, Niels Bohr Institute, University of Copenhagen, 2100 Copenhagen, Denmark}
\affiliation{NNF Quantum Computing Programme, Niels Bohr Institute, University of Copenhagen, 2100 Copenhagen, Denmark}

\author{Valla Fatemi}
\email{vf82@cornell.edu}
\affiliation{School of Applied and Engineering Physics, Cornell University, Ithaca, NY, 14853, USA}

\date{\today}

\begin{abstract} 
Andreev spin qubits are a new qubit platform that merges superconductivity with semiconductor physics.  
The mechanisms dominating observed energy relaxation remain unidentified.
We report here on three steps taken to address these questions in an InAs nanowire weak link. 
First, we designed a microwave readout circuit tuned to be directly sensitive to the spin-dependent inductance of the weak link so that higher orbital states are not necessary for readout -- this resulted in larger windows in parameter space in which the spin state properties can be probed.
Second, we implemented a successful gap-engineering strategy to mitigate quasiparticle poisoning.
Third, the weak link was fabricated by \textit{in situ} shadow evaporation, which has been shown to improve atomic-scale disorder. 
We show how our design allows characterization of the spin stability and coherence over the full range of magnetic flux and gate voltage of an odd parity bias point.
The spin relaxation and dephasing rates are comparable with the best devices previously reported, suggestive that surface atomic-scale disorder and QP poisoning are not linked to spin relaxation in InAs nanowires. 
Our design strategies are transferrable to novel materials platforms for Andreev qubits such as germanium and carbon.
\end{abstract}

\maketitle

\section{Introduction}

Andreev spin states are a new qubit platform composed of microscopic spin degrees of freedom that dictate macroscopic supercurrents~\cite{chtchelkatchev_andreev_2003,park_andreev_2017,tosi_spin-orbit_2019,hays_continuous_2020,hays_coherent_2021,pita-vidal_direct_2023}. 
Such direct integration with superconductivity means naturally strong spin-photon coupling in microwave circuits, and typically experiments intentionally dilute the coupling in order to avoid exiting perturbative regimes~\cite{tosi_spin-orbit_2019,hays_continuous_2020,hays_coherent_2021}. 
On the other hand, near-ultrastrong coupling with a bosonic mode was recently implemented to accomplish long-ranged spin-spin coupling strengths exceeding \SI{200}{\mega\hertz}~\cite{pita-vidal_strong_2024}, demonstrating the potential of this platform for novel quantum computing architectures with long-ranged multi-spin coupling~\cite{pita-vidal_blueprint_2025,lu_kramers-protected_2024}.

A key issue for Andreev spin states is the coherence times seen so far. 
The short dephasing times at the \SI{10}{\nano\second} scale are suspected to arise from the Overhauser field, as the InAs and InSb host materials have a large angular momenta nuclear spin baths~\cite{nadj-perge_spinorbit_2010,hays_coherent_2021,pita-vidal_direct_2023}.
As a result, it will be imperative to develop Andreev spin qubits (ASQs) in materials not hosting dense nuclear spin baths, similar to the transition made for conventional gate-defined quantum dot spin qubits in the past~\cite{de_leon_materials_2021,scappucci_germanium_2021}.
Less understood are the observed spin relaxation times. 
Reported values range from 1 to 50 microseconds~\cite{hays_continuous_2020,hays_coherent_2021,pita-vidal_direct_2023,pita-vidal_strong_2024}, but a systematic investigation has so far been challenging. 
In previous experiments, the ASQs were either strongly entangled with a bosonic degree of freedom~\cite{pita-vidal_direct_2023,pita-vidal_strong_2024}, resulting in ambiguity in the origin of energy relaxation, or experienced strong quasiparticle (QP) poisoning, resulting in limited windows of parameter space to obtain the relaxation time~\cite{hays_continuous_2020}. 
We remark that enhancing relaxation time further can result in ASQs being strongly noise biased, and noise biased qubits can have advantages for error correction implementations~\cite{aliferis_fault-tolerant_2008,webster_reducing_2015,tuckett_ultrahigh_2018}.

Finally, we note that all previous experiments quantifying the spin relaxation time used wires in which the aluminum was etched to create the weak link. 
This process damages the semiconductor material at the atomic scale.
Shadow-evaporated weak links were developed to avoid this damage, leading to lower-disorder devices, as evidenced by both microscopy and quantum transport experiments~\cite{sestoft_shadowed_2024}. 
This sets the stage for a natural question for Andreev spin states: do such minimally-damaged wires exhibit improved spin relaxation? 

Here we probed the spin relaxation of a shadow-evaporated InAs weak link with a microwave readout resonator over a wide range of parameter space. 
Our resonator is designed at an intermediate level of coupling to be directly sensitive to the spin-dependent inductance, avoiding reliance on additional orbital states to accomplish readout. 
We also implemented a superconducting gap engineering strategy inspired by the superconducting qubit community (see, e.g., ~\cite{diamond_distinguishing_2022,connolly_coexistence_2024,glazman_bogoliubov_2021}), and we indeed found QP poisoning to have been mitigated. 
Together, this enabled us to investigate the spin relaxation time over a wide range of gate voltage and flux bias in which the spin is the ground state.
Remarkably, we find no improvement to the spin relaxation time compared to etched wires.
Finally, we investigated the excitation spectra of odd-parity ground states,  finding transitions consistent with ancilla sub-gap states that stabilize phase-dispersive odd-parity states~\cite{sahu_ground-state_2024}, now with measurable spin-orbit effects. 
We speculate that the nuclear spin bath may also be involved in mechanisms that limit the spin relaxation time in existing devices.

\section{Direct spin readout}
\subsection{Nanowire and Resonator design}
\subsubsection{Readout Coupling strength}
\begin{figure}
\centering
\includegraphics[width = 1\columnwidth]{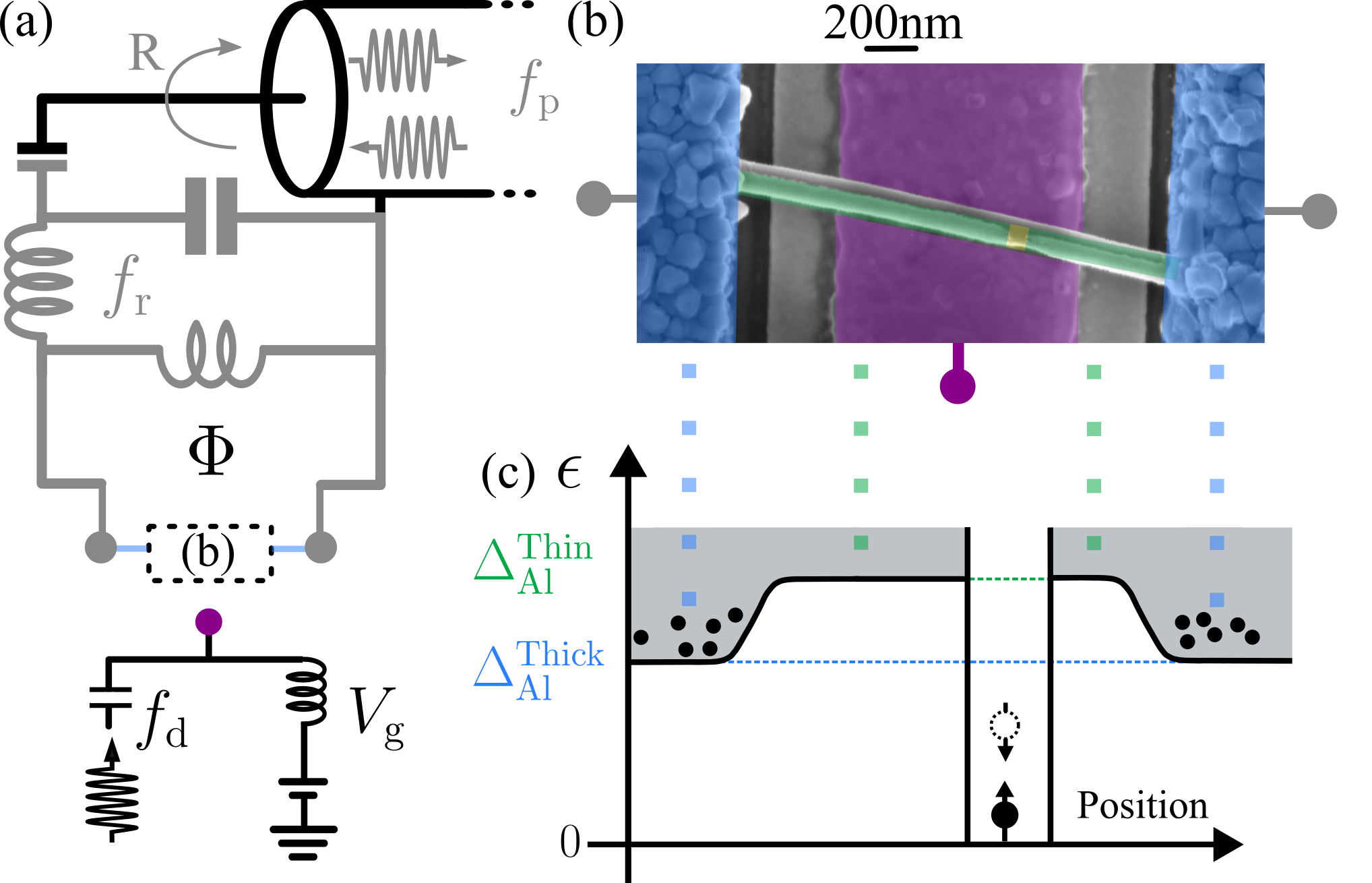}
\caption{ 
\textbf{Device structure and gap engineering strategy.} 
(a) Lumped element circuit diagram of the measured device (details in main text and Appendices~\ref{app:resonator_design} and~\ref{app:fabrication}).
Gray circuit elements represent the readout resonator, $\Phi$ is the external flux applied by a solenoid, and $V_\mathrm{g}$ is the DC gate voltage. $f_\mathrm{p}$ and $f_\mathrm{d}$ represent frequencies incident on the readout resonator and NW gate, respectively.  
(b) False-color scanning electron microscope image of a suspended InAs nanowire with shadow-evaporated weak link and aluminum patch contacts. 
The Al gate electrode is $\SI{100}{\nano\meter}$ thick and Al contact pad thickness is $\SI{200}{\nano\meter}$ thick.
(c) Spatial profile of the superconductor gaps: the epitaxial Al (\SI{25}{\nano\meter} thick) should have a larger gap than the Al resonator (\SI{100}{\nano\meter} thick) and contacting patch (\SI{200}{\nano\meter} thick). 
This increased gap can prevent QPs in the resonator from entering the nanowire (possible ancillary levels not shown). 
}
\label{fig:gapeng} 
\end{figure}

We used a microwave-frequency transmission line resonator that is terminated by a parallel meander inductor (the shared inductance) and nanowire, schematically shown in Fig.~\ref{fig:gapeng}(a).
We designed the shared inductance to be sufficiently large to obtain a direct spin-dependent dispersive shift of order \si{\mega\hertz}.
The spin-dependent contribution to the energy phase relation is typically well-approximated by a sine function
\begin{equation}
    E_\mathrm{s}(\varphi) = E_{\sigma} \sigma_z \sin(\varphi)~,\label{eq:spinEPR}
\end{equation}
where $E_{\sigma}$ is the scale of the spin-dependent energy-phase relation, and $\sigma_z$ is the Pauli $z$ matrix. 
Typically, $E_{\sigma}/h \sim 0.1 - \SI{1}{\giga\hertz}$, corresponding to inductances of several hundred \si{\nano\henry}. 
For our resonator with designed characteristic impedance $Z_\mathrm{c} =\SI{133}{\ohm}$ and effective speed of light $0.39 c$, we found that a shared inductance of about \SI{200}{\pico\henry} is suitable to reliably obtain \si{\mega\hertz}-scale spin-dependent frequency shifts, which are proportional to the second derivative of the spin-dependent energy phase relation Eq.~\eqref{eq:spinEPR}~\cite{metzger_circuit-qed_2021,kurilovich_microwave_2021,fatemi_microwave_2022}. 
The design details are given in Appendix~\ref{app:resonator_design}.

This approach avoids reliance on dispersive interactions involving higher orbital states~\cite{hays_continuous_2020,hays_coherent_2021} while also maintaining a perturbative level of hybridization between the bosonic readout mode and the Andreev states. 
This allows use of shorter weak links in which orbital states are typically at much higher energies, which has the added benefit of higher transparency due to the weak link length being more comparable to the mean free path.

\subsubsection{The shadow-evaporated weak link and quasiparticle gap engineering}
We used InAs nanowires grown by Molecular Beam Epitaxy.
The epitaxial aluminum was deposited \textit{in situ} after the growth of the nanowire, covering three facets of the nanowire. 
The shadowing process resulted in a \SI{136}{\nano\meter} mean length of uncovered nanowire.
The epitaxial aluminum is \SI{25}{\nano\meter} thick.
The nanowire was then placed on top of the resonator electrodes, suspended above a gate electrode below it. 
Note that this design does not introduce intentional gate-based confinement, making our experiment comparable to previous works which relied on poisoning to observe odd parity states~\cite{hays_continuous_2020,hays_coherent_2021,fatemi_microwave_2022,tosi_spin-orbit_2019,sahu_ground-state_2024}. 
Finally, the nanowire was galvanically connected to the resonator by etching the native surface \alox~and immediately depositing a bridging aluminum contact without breaking vacuum.
The resonator and contacting aluminum are \SI{100}{\nano\meter} and \SI{200}{\nano\meter}, respectively.
The final device structure is essentially identical to the one shown in the SEM image in Fig.~\ref{fig:gapeng}(b), although they are different samples.

\begin{figure*}
\centering
\includegraphics[width = \textwidth]{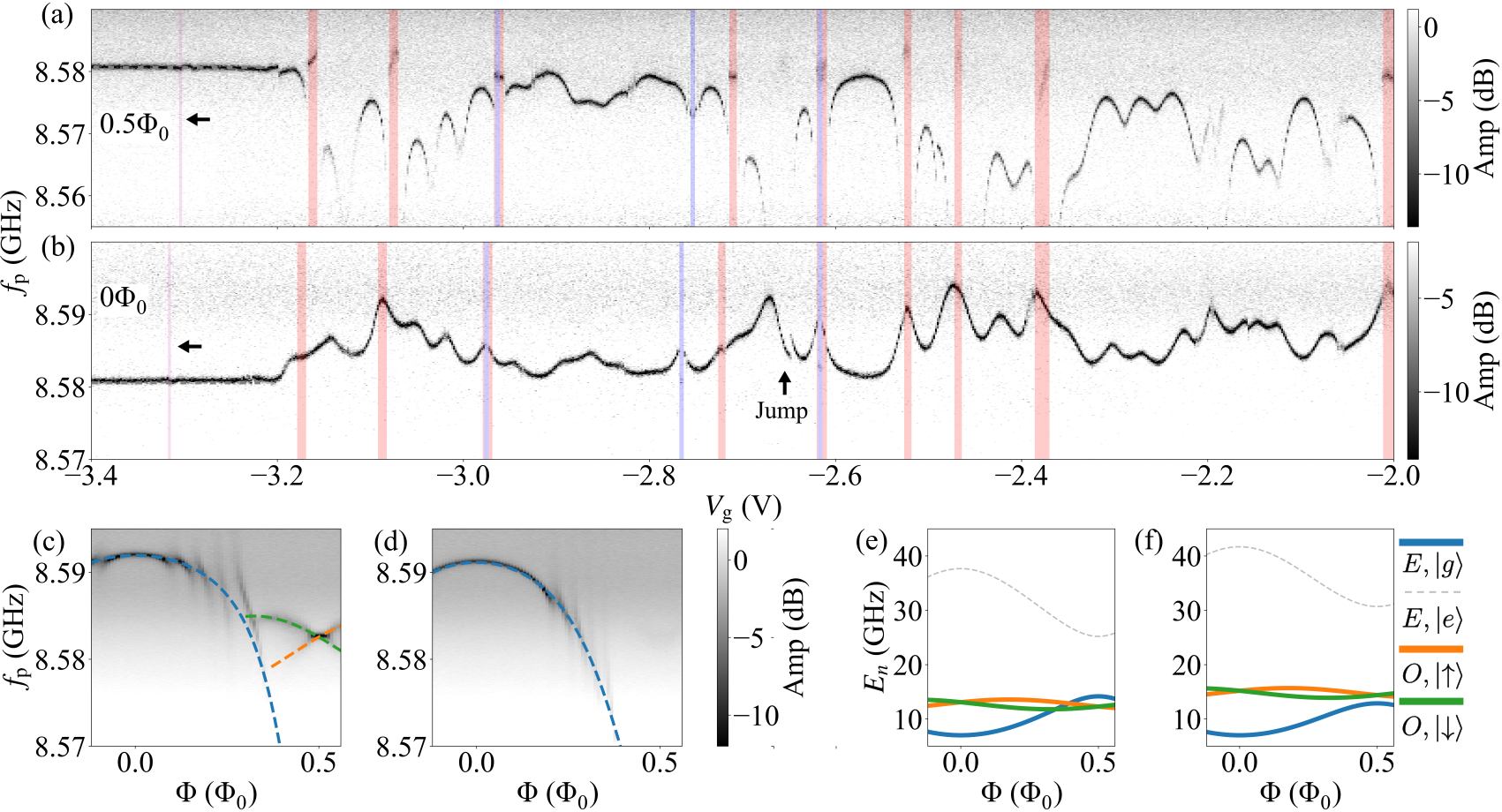}
\caption{ 
\textbf{Single tone measurements and odd-parity stability.} 
Single tone measurement of the resonator while sweeping gate voltage, with mean background subtraction, for (a) half flux, $\Phi = \Phi_0/2$, and (b) zero flux, $\Phi = 0$.
Gate voltage ranges with stable odd parity states are highlighted with colored shading: 
blue indicates stability over the full flux range, while red indicates stability only near $\Phi = 0.5\Phi_0$.
During the $\Phi = 0$ scan, a charge jump occurred near $\SI{-2.7}{\volt}$, as also indicated by a shift of the pinch-off point indicated by the arrows.
Consequentially, the shaded regions for  $V_\mathrm{g}<\SI{-2.7}{\volt}$ in (b) are shifted by $\SI{-12}{\milli\volt}$.
Almost all cases of odd parity stability correspond to a local maximum of resonator frequency corresponding to the even parity state at $\Phi = 0$.
(c)-(d) Single tone resonator measurement while sweeping flux, with mean background subtraction, at $V_\mathrm{g} = \SI{-2.62}{\volt}$, albeit several weeks after (a) and (b) so that some charge drift occurred.
The gate voltage of panel (c) was taken where the odd state is visible for a range of flux, while (d) was taken where the odd state is no longer visible.
The even state was fit with the resonant level model from~\cite{fatemi_microwave_2022} and the odd state is fit with the minimal model from~\cite{padurariu_theoretical_2010}, with extracted parameters: $\Delta_\mathrm{eff}/h = \SI{28.2}{\giga\hertz}, \tau_\mathrm{eff}= 0.85, \it{U}/h = \SI{7}{\giga\hertz}, E_\mathrm{0}/h = \SI{0.4}{\giga\hertz}, E_{\sigma}/h = \SI{0.8}{\giga\hertz}$ and (d): $\Delta_\mathrm{eff}/h = \SI{31.8}{\giga\hertz}, \tau_\mathrm{eff} = 0.7, \it{U}/h = \SI{7}{\giga\hertz}$.
$U$ was inferred from the visibility of odd states between (c) and (d), and we emphasize that it may not reflect the actual charging energy due to ancillary levels shifting the parity transition~\cite{sahu_ground-state_2024} (see main text for details).  
Finally, we remark that the avoided crossings seen in the even state mayb be partially related to higher order couplings to the lowest pair transition (see supplement of ~\cite{fatemi_microwave_2022} for discussion of this effect) or direct resonances with transitions to levels from other channels. 
(e)-(f) Energy-phase relationships (EPRs) using the parameters extracted from the (c) and (d) respectively, assuming the same odd state EPR in (d) as in (c). 
The blue curve represents the lowest even state while orange and green represent the up and down spin states. 
Parity (even, odd) here represents the apparent parity of the weak link -- in later sections, we show that two-tone spectroscopy suggests the presence of an ancillary dot.
}
\label{fig:singletone} 
\end{figure*}

The reason for these material and thickness choices for the superconductors is to provide protection against QP poisoning. 
Here, we are concerned with the fermion parity switches of the Andreev levels hosted inside the weak link due to QPs jumping in from the superconducting reservoirs. 
Because the weak link levels are typically at energies well below the gap edge, if a QP in the electrodes approaches the weak link, it is energetically and thermally favorable for the QP to drop in.
Therefore, a potential barrier is needed to avoid excess QP from reaching the weak link, i.e. to trap the QPs.

The superconducting gap of thin film aluminum has been consistently observed to increase as the thickness decreases~\cite{marchegiani_quasiparticles_2022,cherney_enhancement_1969,court_energy_2007,van_weerdenburg_extreme_2023}.
For conventional tunnel-junction-based devices and qubits, this feature has been used to trap QPs in the lower-gap films to prevent parity switches and QP tunneling relaxation~\cite{aumentado_nonequilibrium_2004,diamond_distinguishing_2022,connolly_coexistence_2024}.
Here, we used this effect to trap QPs in the resonator film as shown schematically in Fig.~\ref{fig:gapeng}(c).
We expect an energy barrier around \SI{13}{\giga\hertz}~\cite{marchegiani_quasiparticles_2022}, sufficient to trap QPs with effective temperature below the \SI{100}{\milli\kelvin} scale (even if the number density may be well out of equilibrium~\cite{connolly_coexistence_2024}).
We also employed best practices to avoid excess non-equilibrium mm-wave and THz photons in the sample space~\cite{serniak_direct_2019,diamond_distinguishing_2022,connolly_coexistence_2024}.

We remark that this structure is distinct from most previous microwave experiments on Andreev levels in Al/InAs-nanowire weak links, which typically used high-gap materials (like Nb or NbTiN) for the resonator~\cite{tosi_spin-orbit_2019,hays_continuous_2020,hays_coherent_2021,pita-vidal_direct_2023,bargerbos_spectroscopy_2023,wesdorp_microwave_2024}. 
When the gap of the resonator is much larger than the epitaxial Al on the nanowire, there may be no local energetic minima to trap nonequilibrium QPs away from the weak link, resulting in poisoning. 
In some of those cases, it was possible to avoid poisoning and stabilize odd-parity ground states by intentionally introducing strong normal confinement, resulting charging energies larger than the aluminum superconducting gap~\cite{bargerbos_singlet-doublet_2022}. 
Recall that here we do not intentionally introduce strong normal confinement. 
Ultimately, we estimate through readout contrast a rough lower bound of the poisoning timescale of \SI{0.7}{\milli\second} (see Appendix~\ref{app:T_poisoning}).
In context of the relatively low transition frequency to the ancillary level, this improves on previous experiments without intentionally strong charging energy by one to two orders of magnitude~\cite{hays_continuous_2020,fatemi_microwave_2022, sahu_ground-state_2024}.

\subsection{Finding stable odd-parity bias points}

We track the readout resonator frequency with standard single-tone microwave reflectometry methods while varying the gate voltage, as shown in Fig.~\ref{fig:singletone}. 
Sequential pairs of discontinuities in the resonator frequency are the first indication for odd-parity ground states.
These are highlighted in Fig.~\ref{fig:singletone} with red and blue shading, where panel (a) shows data taken at half flux quantum and panel (b) shows data taken at zero flux quantum. 
Most of the switches were observed only at half flux quantum (red shading); here, the switch typically occurred as the readout frequency started to diverge due to near-resonant dispersive coupling, consistent with high effective transparency in the lowest energy even-parity states. 
To confirm that these switches represent odd-parity states and not a sudden change of the electric potential of the weak link due to environmental charge jumps, the full flux-dependence confirms a spin-orbit splitting near half-flux quantum (Fig.~\ref{fig:singletone}(c)). 
Overall, about 2\% of the total `active' gate voltage range (from -3.2 to -2\si{\volt}) exhibits a stable odd parity state for for all flux, and 10\% exhibits a stable odd parity state for some flux.

\begin{figure}
\centering
\includegraphics[width = 0.95\columnwidth]{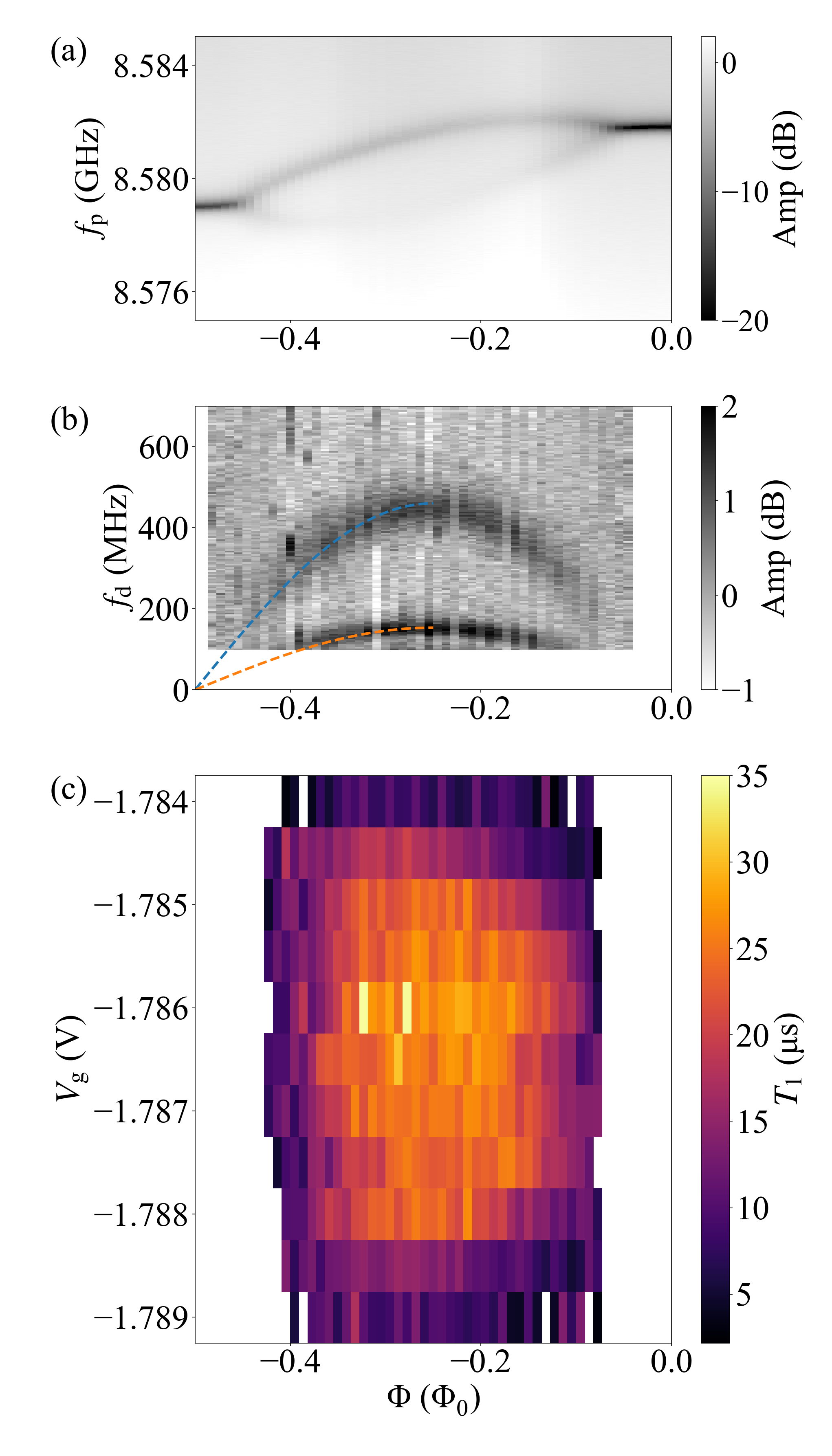}
\caption{ 
\textbf{Spin transition and relaxation time.}
(a) Single-tone resonator measurement at $V_\mathrm{g} = \SI{-1.786}{\volt}$, with mean background subtraction.  
(b) Two-tone spectroscopy, with mean background subtraction, revealing the direct spin-flip transition. 
The blue curve is a minimal model $h f_s(\Phi) = 2E_\sigma\sin{\Phi/\Phi_0}$, where we find $E_\sigma/h = \SI{230}{\mega\hertz}$. 
The orange curve is the blue curve divided by 3, indicating that the lower-frequency signal is a result of higher harmonic generation from the nonlinearity of the bias tee (also verified in room temperature tests of the bias tee). 
(c) The observed relaxation time $T_1$ as a function of gate voltage and flux. 
At this bias point, $T_1$ reaches its maximum value (\SI{30}{\micro\second}) around the middle of the gate and flux stability region ($V_\mathrm{g} = \SI{-1.786}{\volt}$ and $\Phi = -0.25\Phi_0$). 
All panels share the same x-axis.
}
\label{fig:spinT1} 
\end{figure}

These parity switches correlated with local maxima of the frequency shift at zero flux.
This correlation is consistent with local maxima in effective transparency for the lowest energy even-parity states, because the inverse inductance is linearly proportional to transparency in typical models incorporating resonant effects~\cite{kringhoj_anharmonicity_2018,kurilovich_microwave_2021,fatemi_microwave_2022,fatemi_nonlinearity_2024}.
Similarly, the higher stability of odd-parity ground states at half-flux quantum is known in the context of quantum dot Josephson devices ~\cite{bargerbos_singlet-doublet_2022,whiticar_zeeman-driven_2022,van_dam_supercurrent_2006,cleuziou_carbon_2006,jorgensen_critical_2007,jorgensen_openQD_2009,eichler_tuningJJcurrent_2009,lee_zerobias_2012,maurand_first-order_2012,kumar_temperature_2014,szombati_josephson_2016,delagrange_manipulating_2015,delagrange_0pi_2017,garcia_corral_magnetic-field-induced_2020}, although here we do not have an intentionally defined dot. 
We remark that the specific correlation of odd-parity stability at half flux quantum with the local maxima of linear response at zero flux quantum was not pointed out before as far as we know. 

A simple picture can explain this stabilization effect: Fig.~\ref{fig:singletone} (e)-(f)  shows a schematic of the even- and odd-parity states of a resonant level with a small charging energy and spin-orbit interaction. 
For the even-parity states, an effective gap $\Delta_\mathrm{eff}$ and effective transparency $\tau_\mathrm{eff}$ are extracted from Fig.~\ref{fig:singletone} (c)-(d) using a resonant level model~\cite{fatemi_microwave_2022}.
For the odd-parity states, the spin-dependent $E_\mathrm{s}$ and spin-independent $E_\mathrm{0}$ energy of the ASQ are extracted from Fig.~\ref{fig:singletone} (c) using a minimal model~\cite{padurariu_theoretical_2010}.
Near resonance, the effective transparency is high, so a weak charging energy can lower the energy of the odd-parity states enough to result in an odd-parity ground state in a window of flux centered at half flux, as shown in panel (c) and (e). 
As the gate voltage detunes the transparency from the highest value, the odd parity state is disfavored as in panels (d) and (f). 
However, as we will show later, odd-parity stability can also result from the presence of spurious sub-gap levels that can absorb a single QP.

Finally, we remark that the highest local maxima of the resonator frequency at zero flux quantum are around \SI{13}{\mega\hertz} from the bare resonator frequency, which is about $1/6$ of the value one would expect from a simple short junction model at unity transparency. 
This reduction may be due to a combination of unintentional normal confinement (elastic scattering) and the sharp reduction of inductance with the length of the weak link region~\cite{fatemi_nonlinearity_2024}.
Indeed, the \SI{150}{\nano\meter} physical length of the weak link is expected to be comparable to the induced coherence length in the InAs~\cite{williams_proximity_2017,khan_highly_2020}. 
At perfect transparency and assuming a single channel, if the ratio of physical length to the induced coherence length is 1, the inverse inductance of would be $1/3$ of the short-junction value. 
Imperfect transparency would reduce it further and activate transitions into and out of resonances as a function of the chemical potential of the normal region, consistent with the gate voltage dependence we see in the experiment.

For the next sections, we focus on a gate bias point where the ground state was odd-parity over the full flux range, in order to map the properties of the odd-parity state over the full flux range. 

\section{Diagnosis of a stable odd-parity bias point}

\subsection{Dispersive shifts and spin-flip transition}

Here we focus on the odd-parity state near $V_\mathrm{g} = \SI{-1.786}{\volt}$ in Fig.~\ref{fig:singletone}(c). 
Fig.~\ref{fig:spinT1}(a) shows the flux-dependence of the readout resonator in the odd-stable region, showing clear spin-orbit split states. 
The strength of the spin-orbit splitting in the readout resonator, given our designed sensitivity to inductance, is suggestive of $E_{\sigma}/h \approx \SI{255}{\mega\hertz}$ (see Appendix~\ref{app:fr_EPR_main} for details of this conversion).

We compare this directly with the spin-flip transition frequency. 
We conducted two-tone spectroscopy experiments on the weak link by driving with a short \SI{74}{\nano\second} pulse on the gate electrode followed by a readout pulse. 
We observed direct spin-flip transitions, which has been reported as an electron dipole spin resonance (EDSR)~\cite{metzger_circuit-qed_2021,pita-vidal_direct_2023,bargerbos_spectroscopy_2023}.
The dotted blue line in Fig.~\ref{fig:spinT1}(b) follows the spin-flip transition as a function of flux assuming a pure sinusoid, per Eq.~\ref{eq:spinEPR}.
We find reasonable agreement with this simple function for $E_{\sigma}/h \approx \SI{240}{\mega\hertz}$, within 10\% of the value estimated from the readout frequency shift. 
This close agreement validates the intention of our resonator design.
We remark that the spin-flip transition remains visible down to $\SI{250}{\mega\hertz}$, suggesting that the spin is thermalized down to near $\SI{15}{\milli\kelvin}$.

\subsection{Spin relaxation rate}

With spin-dependent readout and the spin-flip transition frequency established, we turn to assessments of the spin relaxation time. 
The spin lifetime at different flux was previously reported over a narrow range of flux $\pm 0.1\Phi_0$~\cite{hays_continuous_2020}.
To our knowledge, a full mapping of the relaxation rate as a function of flux and gate has not been reported.

We drove the spin transition with a Gaussian pulse (FWHM = $\SI{74}{\nano\second}$), and then we checked for spin relaxation by a readout pulse following a variable delay~\footnote{We found that a long saturation pulse may result in double-exponential decay -- we show this in Appendix~\ref{app:T1meas}.}.
We found a single-exponential decay, from which we extract a decay time $T_1$ (see Appendix~\ref{app:T1meas}). 
The two-dimensional scan of $T_1$ as a function of flux and gate voltage is shown in Fig.~\ref{fig:spinT1}(c) for this bias point. 
The boundaries of the map are either where the spins were indistinguishable in readout or the lifetime was too short.
We found that $T_1$ reaches maximum in the middle of the scanning range, roughly centered in both flux (near $\Phi_0/4$) and gate voltage (in the middle of the odd-parity stable range). 
This maximum value near \SI{30}{\micro\second} is comparable to the highest values reported previously in the literature for Andreev spin states in Al/InAs nanowires~\cite{hays_continuous_2020,hays_coherent_2021}. 
Measurements of $T_1$ with flux at different gate voltage bias points are shown in Appendix~\ref{app:extra_bias}.

\subsection{Higher frequency transitions}\label{chp:Higher_ft}

\begin{figure*}
\centering
\includegraphics[width = 1\textwidth]{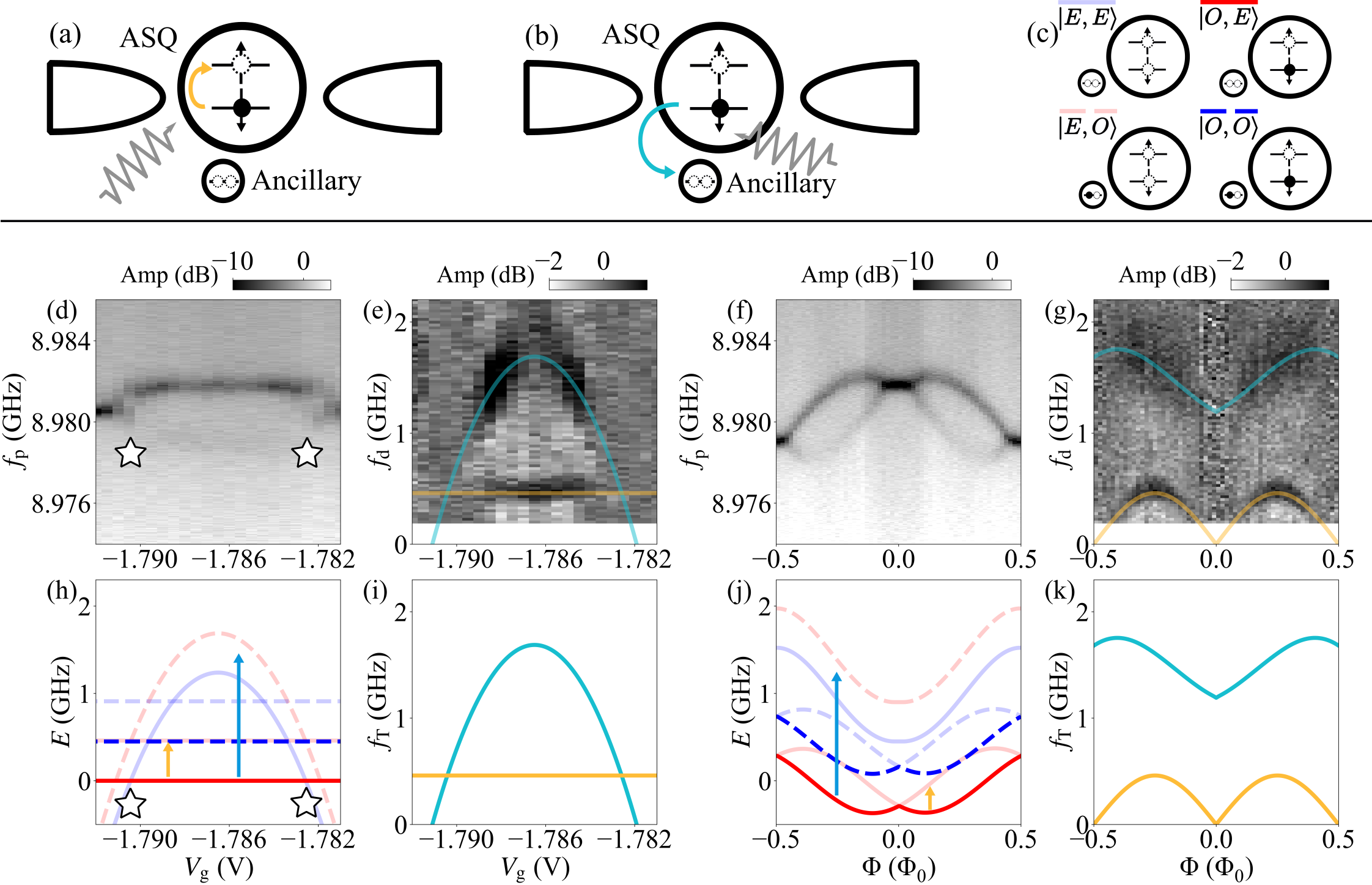}
\caption{ 
\textbf{Possible parity flipping transitions at higher frequency than the spin-flip transition.}
(a)-(c) Schematic of possible transitions and QP configurations. 
(a) depicts the spin-flip transition, while (b) depicts a possible parity-flipping transition involving transfer of a QP between the Andreev level and an ancillary dot state, as proposed by Sahu, et al.~\cite{sahu_ground-state_2024}. 
(c) depicts four possible occupations of QPs involving the ancillary dot, serving as a legend for later plots. 
Labels in the kets represent the parity of the weak link and the ancillary dot, respectively ($E$ for even, $O$ for odd). 
(d) Gate dependence of single-tone spectroscopy at $\Phi = -0.25\Phi_0$. 
Between $V_\mathrm{g} = \SI{-1.790}{\volt}$ and $V_\mathrm{g} = \SI{-1.782}{\volt}$ we observe two readout peaks (one weak), indicating the weak link has odd-parity ground state. 
Stars in (d), (h) indicate the even-odd switching gate bias points predicted by the model in (h).
(e) Gate dependence of two-tone spectroscopy at $\Phi = -0.25\Phi_0$.  
The non-dispersing transition near $\SI{0.5}{\giga\hertz}$ is the spin-flip transition.
The higher frequency dispersing transition, depicted schematically in (b), is fit using model explained in Sec.~\ref{chp:Higher_ft}.
(f) Flux dependence of single-tone resonator spectroscopy at $V_\mathrm{g} = \SI{-1.7855}{\volt}$.
(g) Flux dependence of two-tone spectroscopy at $\SI{-1.7855}{\volt}$.
Line colors follow the same conventions and using a consistent set of parameters across panels (d-g). 
(h) Gate dependence of the energy of four different parity states in (c) (with $|O,E\rangle$ and $|O,O\rangle$, each having two spin manifolds, presented in normal and dim color), with parameters extracted from (d)-(g):
$E_\mathrm{0}/h = \SI{291}{\mega\hertz}$ and $E_\sigma/h = \SI{230}{\mega\hertz}$. 
$\epsilon_\mathrm{qd}$ can not be extracted independently from $U$, so we choose $\epsilon_\mathrm{qd} = \SI{450}{\mega\hertz}$ for demonstration. 
The dim solid red line and dim dashed blue line represent the higher energy spin state of the globally-odd manifold.
(i) Possible transition of state $|O,E\rangle$ (solid red in (h)) to other states of the same global parity. 
(j) EPRs of the four different parity states in (c).
(k) Possible transition of state $|O,E\rangle$ (solid red in (h)) to other states of the same global parity. 
}
\label{fig:TT_higher} 
\end{figure*}

To investigate the possible source of sensitivity of $T_1$ to gate voltage, we inspected higher frequencies in two-tone spectroscopy, shown in Fig.~\ref{fig:spinT1}. 
The spin-flip transition was relatively gate-voltage independent, but a higher frequency transition that was strongly gate-voltage dependent is visible.
This was the only other transition seen between 0 and 10 GHz at this bias point (Appendix~\ref{app:higher_fre_TT}).

This transition frequency peaks at the center of the odd-parity stable region ($V_\mathrm{g}=\SI{-1.7865}{\volt}$) and trends toward zero where the parity transitions occur. 
This aspect is consistent with recent proposals for parity flipping as a result of ancillary dot-like levels~\cite{sahu_ground-state_2024}, and suggests that the transition was not to higher levels within the odd-parity manifold~\cite{tosi_spin-orbit_2019,hays_continuous_2020,matute-canadas_signatures_2022}.
The lifetime of this excited state was too short for us to measure, and we only detected it when it mediates a spin flip after the QP relaxes back to the ground state.
This suggests that the ancillary level is hosted in the semiconductor in a region with a distinct spin-orbit flavor relative to the main Andreev level, rather than via a decoupled droplet of aluminum.

The ancillary level interpretation implies that microwaves can induce apparent parity flips by transitioning a QP between a strongly flux-dispersing Andreev level and a weakly flux-dispersing dot-like level, as discussed in depth by Sahu, et al.~\cite{sahu_ground-state_2024}. 
The minimal model is of a gate independent energy of the ancillary state $E_\mathrm{qd} = n_d \epsilon_\mathrm{d}$ and energy of the ABS $E_\mathrm{A} = (n_\mathrm{A}-1)\epsilon_\mathrm{A} + U(n_\mathrm{A}-1)^2$, where $n_d$ ($n_\mathrm{A}$) refers to the QP occupation of the ancillary state (Andreev level) and $\epsilon_d$ ($\epsilon_\mathrm{A}$) refers to the energy of the ancillary state (Andreev level).
We modeled the gate dependences as $\epsilon_\mathrm{A}(V_\mathrm{g}) = \epsilon_\mathrm{A}(V_\mathrm{g}^\mathrm{c})+\alpha (V_\mathrm{g}-V_\mathrm{g}^\mathrm{c})^2$ when $V_\mathrm{g}$ is in the vicinity of gate voltage at the center of the stability $V_\mathrm{g}^\mathrm{c}$, coinciding with where the weak link has maximum local transparency.
Consequentially, the transition frequency of this mechanism is $U+\epsilon_\mathrm{qd}-\epsilon_\mathrm{A}(V_\mathrm{g})$, similar to the parabolic line shape we observed here.

The higher transition frequency dispersed strongly with flux, which includes contributions from the flux-dependence of both the odd- and even-parity Andreev states (within the same global parity manifold). 
We compared the transition frequency with the readout frequency shifts of the odd states and the nearby regions of stable even-parity states. 
The corresponding transition is shown in Fig.~\ref{fig:TT_higher} (e) and (g).
The flux- and gate-dependent part of the transition has the form
\begin{equation}
    U+ \epsilon_\mathrm{qd} -\epsilon_\mathrm{A}(\varphi, V_\mathrm{g})-E_\mathrm{0}\cos(\varphi) + |E_{\sigma}\sin(\phi)| 
\end{equation}
where $\epsilon_\mathrm{A}$ represents Andreev level, and $E_\mathrm{0} = \SI{-291}{\mega\hertz}$ and $E_{\sigma} = \SI{230}{\mega\hertz}$ represent the spin-independent and spin-dependent coefficient of the ASQ EPR.
Our experiment cannot quantify the phase-independent part of the first three terms, $U+\epsilon_\mathrm{qd}-\epsilon_\mathrm{A}(\phi, V_\mathrm{g})$ individually.
From two-tone spectroscopy, we extract $U+\epsilon_\mathrm{qd}-\epsilon_\mathrm{A}(0, \SI{-1.755}{\volt})=\SI{900}{\mega\hertz}$, and $\alpha=\SI{80}{\mega\hertz~\milli\volt^{-2}}$.
Values of the phase-dependent parts of $\epsilon_\mathrm{A}(\varphi),E_\mathrm{s}$ are independently extracted from single-tone readout spectroscopy (Appendix~\ref{app:fr_EPR_even},\ref{app:fr_EPR_main}).
As shown with the cyan curves in Fig.~\ref{fig:TT_higher} (g), this model captures the flux and gate dependent feature of the anomalous transition.

\section{Discussion}

Our data shows largely even-parity ground states at zero flux, and no weakly-coupled ``dotty" region with strong charging energy at gate voltages before the activation of the nanowire inductance. 
This is consistent with shadow-evaporated weak links exhibiting improved transparency relative to etched weak links~\cite{sestoft_shadowed_2024}. 
However, we did not find any improvement in the spin relaxation time relative to previous work, suggesting that the atomic-scale disorder induced by the etch process in previous experiments does not induce new spin relaxation mechanisms. 

At the bias point presented in the main text, our data are consistent with the trend seen by Hays, et al.~\cite{hays_continuous_2020}, with increasing spin relaxation time as magnetic flux detunes from the half-flux quantum Kramers point. 
In this experiment we found the same trend with flux around both Kramers points (zero and half flux quantum), although with differences in the details of the flux-dependence. 
At other bias points with spin transition frequencies higher than here or in Hays, et al., the flux-dependence of the relaxation rate is weaker (see Appendix~\ref{app:extra_bias}). 

Nonetheless, all of these trends with the spin transition frequency are inconsistent or opposite to that predicted from $1/f$ charge or flux noise~\cite{padurariu_theoretical_2010} as well as for phonon-mediated spin relaxation~\cite{erlingsson_hyperfine-mediated_2002,fujisawa_spontaneous_1998,golovach_phonon-induced_2004,trif_spin_2008,kornich_phonon-mediated_2014,ungerer_coherence_2024}.
Hyperfine interactions are thought to limit dephasing, and we remark that there is potentially a connection between dephasing and relaxation. 
Specifically, at Kramers' point, the spin states are energetically degenerate, so that relaxation and dephasing no longer have distinct meaning -- hyperfine interactions would induce random spin splittings and rotations. 
As flux detunes from Kramers' point, relaxation and dephasing gain distinct meaning, and $T_1$ might be expected to increase as the nuclear spins are unable to directly absorb the energy from the Andreev spin. 
As the spin-splitting continues to increase, a progressively increasing density of states of phonons would mediate spin relaxation, resulting in a decreasing trend of $T_1$~\cite{erlingsson_hyperfine-mediated_2002,fujisawa_spontaneous_1998,golovach_phonon-induced_2004,trif_spin_2008,kornich_phonon-mediated_2014,ungerer_coherence_2024}.
However, that the crossover could be around or above \SI{1}{\giga\hertz} remains to be explained.

Finally, we remark that our design considerations for the readout resonator, weak link design, and gap engineering are straightforwardly extended to new materials platforms for Andreev states. 
Such a probe will help to probe spin relaxation and dephasing over wide parameter ranges while avoiding QP poisoning.

\section*{Data and code availability}
All data generated and code used in this work are available at: \textcolor{black}10.5281/zenodo.14628787.

\begin{acknowledgments}
\textit{Personal Acknowledgements}
We acknowledge assistance from Pavel Kurilovich and Thomas Connolly on the microwave packaging and resonator design, Wei Dai and Michel Devoret for the fast microwave switch, Joachim Sestoft and Rasmus Schlosser on shadow junction development, discussion with Kushagra Aggarwal regarding higher frequency transitions, and discussions with Stefano Bosco, Leonid Glazman, Srijit Goswami, and Pavel Kurilovich regarding spin and QP relaxation times.

V. F. and M. K. acknowledge the LPS Quantum Computing Program Review for establishing the connection leading to this collaboration. 

\textit{Funding Acknowledgements}
Research by V.F. and H. L. was sponsored by the Army Research Office and was accomplished under Grant W911NF2210053 Number W911NF-22-1-0053. The views and conclusions contained in this document are those of the authors and should not be interpreted as representing the official policies, either expressed or implied, of the Army Research Office or the U.S. Government. The U.S. Government is authorized to reproduce and distribute reprints for Government purposes notwithstanding any copyright notation herein.

M. K., D. F. B. and Z. S. gratefully acknowledge support U.S. Army Research Office Grant No. W911NF-22-1-0042, Villum Foundation (grant 37467) through a Villum Young Investigator grant, the European Union (ERC Starting Grant, NovADePro, 101077479), and the Novo Nordisk Foundation, Grant number NNF22SA0081175, the NNF Quantum Computing Programme. 
T. K. and J. N. acknowledge support from the Danish National Research Foundation (DNRF 101), European Union’s Horizon 2020 research and innovation programme under grant agreement FETOpen grant no. 828948 (AndQC), the Novo Nordisk Foundation project SolidQ and the Carlsberg Foundation. 
Any opinions, findings, conclusions or recommendations expressed in this material are those of the author(s) and do not necessarily reflect the views of Army Research Office or the US Government nor do they necessarily reflect those of the European Union or the European Research Council. Neither the European Union nor the granting authority can be held responsible for them. 
\end{acknowledgments}

\section*{Author Contributions}
H. L. designed the device, acquired the data, and analyzed the data under the supervision of V. F..
T. K. and J. N. developed the nanowires with shadow junctions.
Z. S. and D. B. F. fabricated the device under guidance by M. K. 
V. F. conceived the experiment and established the collaboration with M. K.. 
H. L. and V. F. wrote the manuscript with input from all authors. 

\section*{Competing Interests}

The authors declare no competing interests.

\clearpage
\onecolumngrid

\renewcommand{\thefigure}{S\arabic{figure}}
\renewcommand{\thetable}{S\arabic{table}}

\setcounter{figure}{0}
\renewcommand{\thefigure}{S\arabic{figure}}
\renewcommand{\theHfigure}{S\arabic{figure}}

\appendix

\begin{titlepage}
  \centering
  \vskip 60pt
  \LARGE \textbf{Appendices: Andreev spin relaxation time in a shadow-evaporated InAs weak link} \par 
  \par
  \vskip 2em
\end{titlepage}
  
\maketitle

\begin{figure*}
\centering
\includegraphics[width = \textwidth]{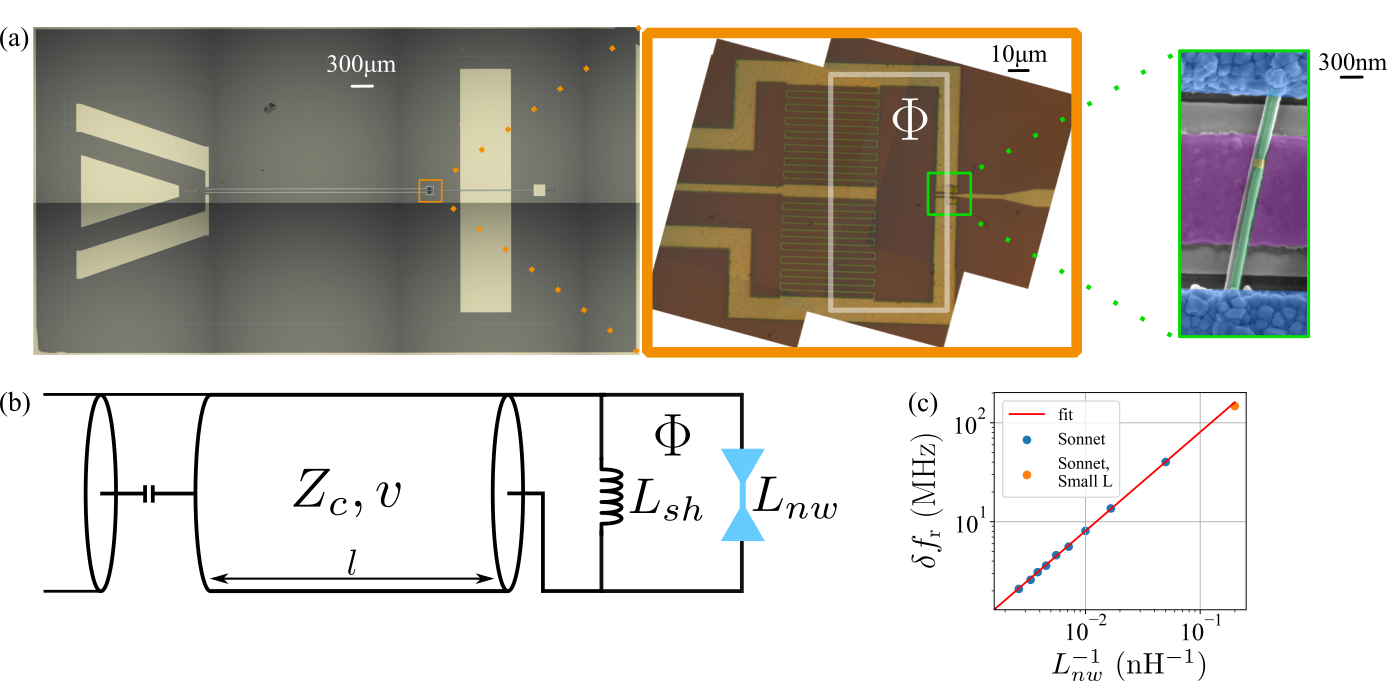}
\caption{\textbf{Distributed circuit diagram and readout frequency sensitivity.}
(a) OM and SEM images showing the general and detailed device structure.
(b) Schematic representing the distributed-element resonator and coupling structure used in this experiment.
(c) Resonator frequency sensitivity to nanowire inductance determined from finite element simulations in Sonnet Suites\textsuperscript{TM}.
Bare resonator frequency in simulation $f_\text{r}^\text{sim}=\SI{8.57}{\giga\hertz}$, with fitted slope $S = \SI{803}{\mega\hertz\nano\henry}$.
The orange dot drops slightly out of the linear fitting line as it is no longer in the ideal dispersive regime. 
}
\label{fig:res} 
\end{figure*}

\section{Resonator design} \label{app:resonator_design}

In our experiment, we implement a distributed element coplanar strip resonator that couples the nanowire to the resonator's differential mode via the nanowire inverse inductance, as depicted in Fig.~\ref{fig:res}(a)-(b). 
The nanowire inverse inductance is proportional to the second derivative of the energy-phase relation~\cite{metzger_circuit-qed_2021,kurilovich_microwave_2021, fatemi_microwave_2022}.
The coupling inductance is designed so that the resonator has a sensitivity $S = \SI{803}{\mega\hertz~\nano\henry}$ to the nanowire inverse inductance (Fig.~\ref{fig:res}(c)).
This value is targeted so that the resonator frequency will achieve spin-dependent frequency shifts that are useful for readout due only due to the spin-dependent part of the energy-phase relation. 
Additional design parameters are the coupling rate to the feedline \SI{0.3}{\mega\hertz}, the characteristic resonator impedance $Z_\mathrm{c} = \SI{133}{\ohm}$, and the light velocity $v= 0.39 c$.

\section{Device Fabrication} \label{app:fabrication}

The device is fabricated on high resistivity silicon ($\SI{20}{\kilo\ohm\cdot\centi\meter}$). 
The wafer is dipped into buffered hydrofluoric (BHF) Acid to remove the native silicon oxide ($\text{SiO}_2$) prior to aluminum (Al) deposition. 
The wafer is then immediately loaded into the molecule beam epitaxy (MBE) system. 
After degassing overnight, \SI{100}{\nano\meter} Al is deposited in the metal deposition chamber (MDC) of the system. 
The wafer is diced into $\SI{10}{\milli\meter}\times\SI{10}{\milli\meter}$ chips protected with photo resist.
The protection photo resist is stripped with 1,3-dioxolane, acetone and isopropyl alcohol (IPA). Ultrasonicaton is used to help with the clean procedure. 
A layer of PMMA A4 e-beam resist is spun coated and then baked at \SI{185}{\celsius} for \SI{2}{\minute}. 
The fine features of the design are exposed with \SI{3}{\nano\ampere} beam current in the Elionix ELS-F125 e-beam lithography system. 
The pattern is developed with a solution of MIBK:IPA 1:3 and wet etched with Transene D aluminum etchant. 
Then, the resist is stripped in \SI{50}{\celsius} acetone and later oxygen plasma ash. 
A similar procedure is performed to pattern the coarse features of the design, with the main difference being a higher exposure current of \SI{60}{\nano\ampere}. 

The gate line is fabricated through lift-off procedure.
Before the Al of the gate is deposited, \SI{15}{\nano\meter} Hafnium Oxide ($\text{HfO}_2$) dielectric layer is grown with atomic layer deposition (ALD) to prevent gate leakage. 
The gate is then evaporated in the e-beam evaporation system from AJA international, Inc. Lift-off is performed in hot acetone with low power ultrasonication. 
The combined thickness of the $\text{HfO}_2$ and the Al is less than \SI{100}{\nano\meter} to ensure spacing between the nanowire and the gate line. 
MBE grown half shell InAs/Al nanowires (batch number QDev1018~\cite{schlosser_shadow_2021}) are deterministically transferred from the growth substrate to the device with a micro-manipulator machine. 

The nanowires have weak links defined by shadows from other nanowires during the low temperature Al deposition~\cite{sestoft_shadowed_2024}. 
Since the shadow junction growth is not deterministic, the nanowire junctions are checked with scanning electron microscopy (SEM). 
Nanowire position adjustment and SEM checks are performed until the nanowire shadow junction lies on top of the gate line.
Then, to ensure good galvanic connection between the nanowire and the control layer, the native Al oxide of the nanowire and the control layer is removed by argon milling inside the AJA. Then \SI{200}{\nano\meter} of Al are evaporated to ensure the connection.

\section{Experimental Setup}\label{app:expt_setup}

The device was measured in a Bluefors dilution fridge with $\approx \SI{10}{\milli\kelvin}$ base temperature. 
The RF cable setup contains one readout loop, one spin drive channel and one TWPA pump channel. 
The DC wire setup contains one gate channel for controlling the doping of the NW and one flux channel for controlling the solenoid, which is inductively coupled to the device. 
The input channel has $\SI{60}{\deci\bel}$ attenuation from multiple attenuators located at the interface of stages and $\approx \SI{7}{\deci\bel}$ from coax cables, one KL filter and multiple Eccosorb filters. 
The device is placed within an aluminum shield can inside the MXC chamber, which further suppresses flux noise from the surrounding environment and stray photons. 
The output channel has one TWPA with $\approx \SI{30}{\deci\bel}$ gain at MXC chamber, one three-terminal isolator, one KL filter, one HEMT with $\approx \SI{37}{\deci\bel}$ gain at 4K stage and one room-temperature amplifier with $\approx \SI{38}{\deci\bel}$ gain. 
The TWPA is pumped via the TWPA pump channel of $\SI{36}{\deci\bel}$ attenuation from attenuaors and one $\SI{20}{\deci\bel}$ directional coupler.
The DC branch of the gate channel is filtered by RC+RF Qdevil filters, combined with a spin drive signal via a bias-tee located at MXC chamber. 
The flux bias line is filtered by Thor lab EF508 and EF120 low pass filters at room temperature.

\begin{figure*}
\centering
\includegraphics[width = \textwidth]{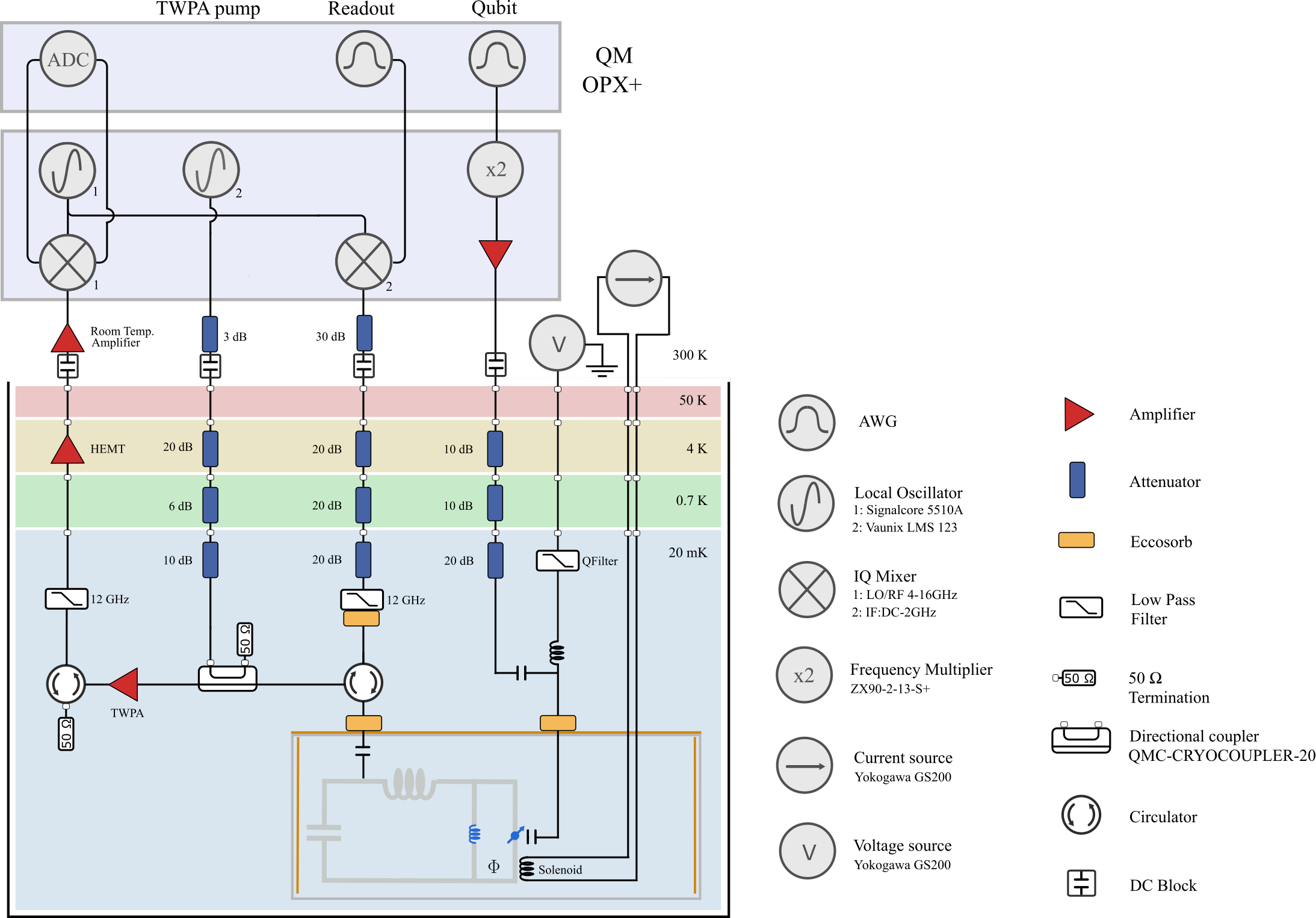}
\caption{\textbf{Experimental wiring diagram.} Schematic of electronics inside dilution fridge and at room temperature.
}
\label{fig:wire} 
\end{figure*}

The input and output is generated and demodulated by Quantum Machine OPX+ system, frequency range within $\SI{400}{\mega\hertz}$. 
A homemade interferometer consist of one local oscillator (LO) with $8-\SI{12}{\giga\hertz}$ output range, one IQ mixer up-converting the signal and one IF mixer down-converting the signal. 
The TWPA pump is powered directly by an LO. 
The spin drive used in this experiment is through a $2\times$ frequency multiplier to achieve a frequency range of $0-\SI{800}{\mega\hertz}$, and amplified by RT amplifier to compensate for conversion loss. 
The details of the setup are shown in Fig.~\ref{fig:wire}.

\section{Extracting Even-parity Andreev bound state parameters}\label{app:fr_EPR_even}

\begin{figure}
\centering
\includegraphics[width = 1\columnwidth]{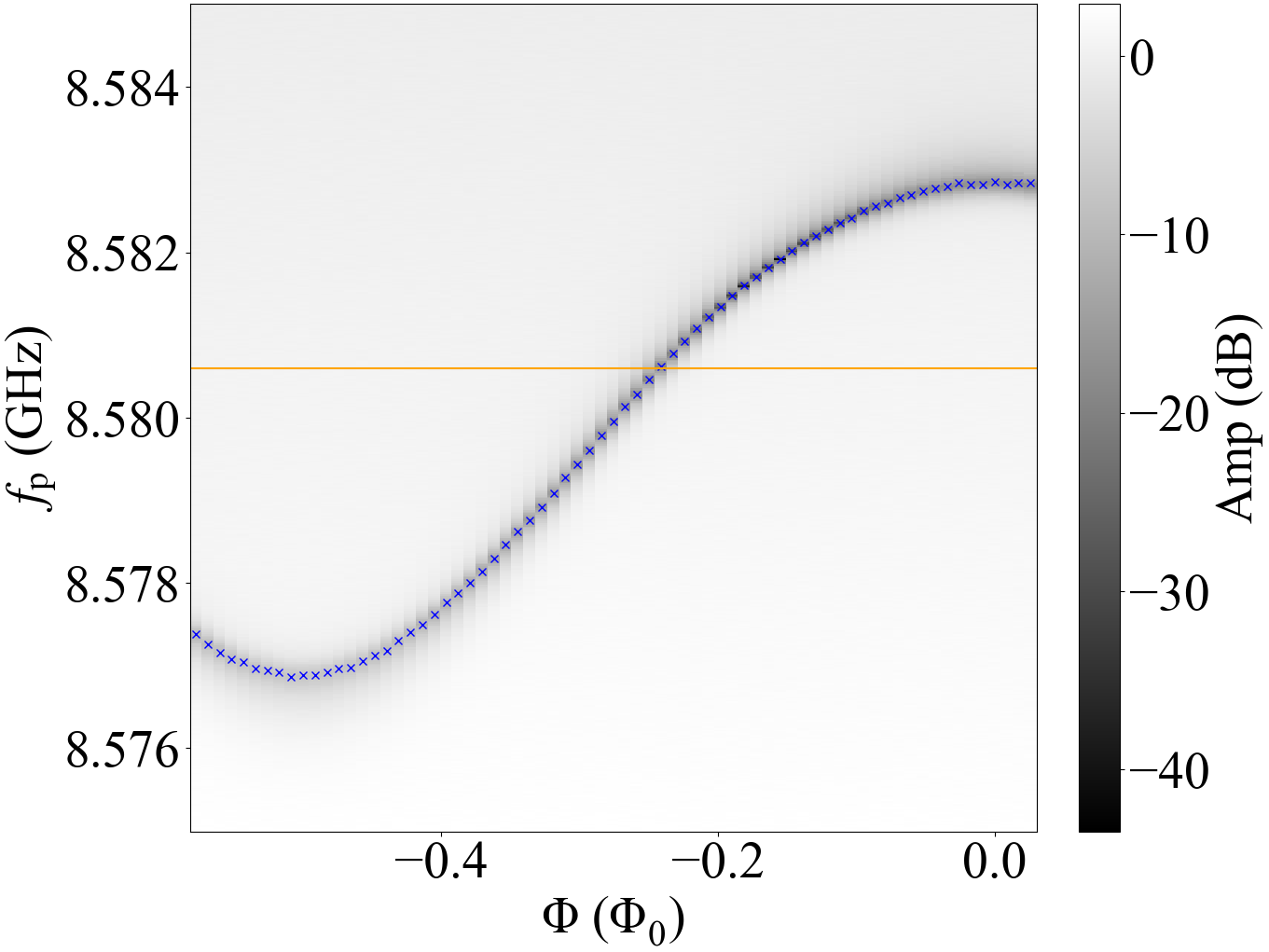}
\caption{\textbf{Obtaining EPR of even-parity manifold from single-tone readout spectroscopy.}
Single tone readout at $V_\mathrm{g} = \SI{-1.775}{\volt}$, near singlet-doublet switch boundary. 
The orange line is the bare resonator frequency $f_r = \SI{8.5806}{\giga\hertz}$. 
The blue crosses are the detected resonator frequency at different flux.}
\label{fig:SI_even_singletone} 
\end{figure}

In Fig.~\ref{fig:singletone} and Fig.~\ref{fig:TT_higher}, we used EPR of the even-parity manifold extracted from single-tone readout spectroscopy.

In Fig.~\ref{fig:TT_higher}, we first extract readout frequency versus flux, as shown in Fig.~\ref{fig:SI_even_singletone}.
The relationship between inverse inductance and EPR of the weak link can be derived as
\begin{equation}
    \frac{1}{L_\mathrm{NW}} = \left(\frac{2\pi}{\Phi_0}\right)^2\frac{\partial^2E}{\partial\varphi^2}~,
\end{equation}
with $\varphi$ the phase across the weak link.

As introduced in Appendix~\ref{app:resonator_design}, the relationship between frequency shift and EPR in the dispersive regime can be interpreted as
\begin{equation}
    \delta f(\varphi) = S L_\mathrm{NW}^{-1} = S\left(\frac{2\pi}{\Phi_0}\right)^2\frac{\partial^2E}{\partial\varphi^2}~.
    \label{eq:EPR_singletone}
\end{equation}

Due to time-reversal symmetry, $E'(n\pi)  = 0, n\in \mathbb{Z}$~\cite{titov_negative_2001}.
The EPR of the even-parity manifold can be calculated from the frequency shift using:
\begin{align}
    E'(\varphi) &=  \left(\frac{\Phi_0}{2\pi}\right)^2S^{-1} \int_{0}^{\varphi} \delta f(\varphi'') d\varphi''~,\\
    E(\varphi) - E(0) &= \int_{0}^{\varphi}E'(\varphi')d\varphi~,
\end{align}
in our code, we conduct discrete integration using Simpson's rule.

In Fig.~\ref{fig:singletone}, we fit the single-tone readout spectroscopy using an EPR derived from resonant level model~\cite{beenakker_resonant_1992,kurilovich_microwave_2021,fatemi_microwave_2022} and Eq.~\ref{eq:EPR_singletone}.
Here we outline the calculation of the EPR and its derivatives for the resonant level model as described in~\cite{fatemi_nonlinearity_2024}.
First, we have the scattering amplitude, which is defined by
\begin{equation}
    \begin{split}
        \Lambda\left(\epsilon, \varphi\right)&=-2 i \Gamma \epsilon^2 \sqrt{\epsilon^2-\Delta^2}+\left(\epsilon_0^2-\epsilon^2+\Gamma^2\right)\left(\epsilon^2-\Delta^2\right)+\\
        &\Delta^2\left(\Gamma^2-\delta \Gamma^2\right) \sin ^2\left(\frac{\varphi}{2}\right)~,
    \end{split}
\end{equation}
with $\Gamma$ the overall coupling energy of the weak link to the lead, $\delta \Gamma$ the difference of the coupling energy between the weak link to the left and right lead, $\epsilon_0$ dot offset energy, and $\varphi$ the static phase across the weak link.
Effective gap $\Delta_{\text {eff }}=\Delta \sqrt{\Gamma^2+\epsilon_0^2}/(\Delta+\Gamma)$ and effective transparency $\tau_{\text{eff}}=(\Gamma^2-\delta \Gamma^2)(\Gamma^2+\epsilon_0^2)$.

We can then write the scattering amplitude into energy-dependent and phase-dependent terms, respectively. 
Defining $\bar{\Lambda}(i \epsilon)$ as the energy-dependent term normalized by the amplitude of the phase-dependent term:
\begin{equation}
    \begin{split}
        \Lambda\left(\epsilon, \varphi\right)&=\Lambda_{\epsilon}(\epsilon)+\Lambda_{\varphi}(\varphi)~,\\
        \bar{\Lambda}(i \epsilon)&=\Lambda_{\epsilon}(i \epsilon)/2\Lambda_{\varphi}(\pi)~.
    \end{split}
\end{equation}
With this, the ground state EPR can be written in the form
\begin{equation}
U(\varphi)=-\frac{1}{2 \pi} \int_{-\infty}^{\infty} \log \left(1+\frac{1-\cos (\varphi)}{\bar{\Lambda}(i \epsilon)}\right) d \epsilon~.
\end{equation}
To solve for the inverse inductance at $\varphi=\varphi_0$, we can take a second derivative of the argument to the integral
\begin{equation}
c_2(\varphi_0)=-\frac{1}{2 \pi 2!} \int_{-\infty}^{\infty} \frac{\partial^2}{\partial \varphi^2} \log \left(1+\frac{1-\cos (\varphi)}{\bar{\Lambda}(i \epsilon)}\right)\biggr\rvert_{\varphi=\varphi_0} d \epsilon~,
\end{equation}
where $2c_2$ is the inverse inductance.
In Fig.~\ref{fig:singletone}, the even-parity ground state is fit by Eq.~\ref{eq:EPR_singletone} using $2c_2$ as inverse inductance.
We remark that this formulation includes continuum contributions to the EPR.

\section{Extraction of spin-flip transition frequency}

\subsection{Single-tone spectroscopy}\label{app:fr_EPR_main}

\begin{figure}
\centering
\includegraphics[width = 1\columnwidth]{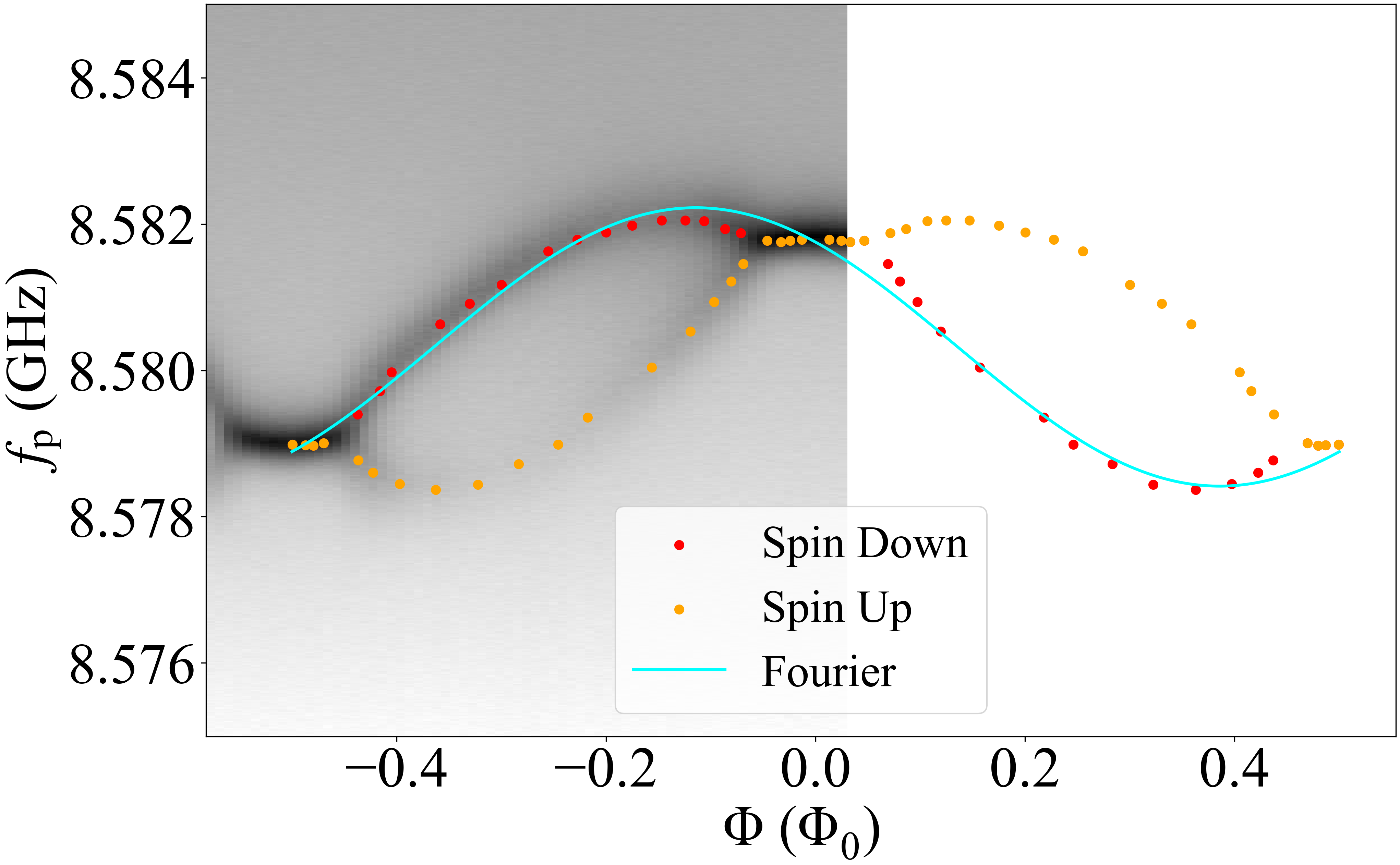}
\caption{ 
\textbf{Fit single-tone spectroscopy of the readout resonator.}
Red dots: readout frequency of spin down branch at different flux.
Orange dots: readout frequency of spin-up branch at different flux.
Cyan curve: the fundamental frequencies of the cosine and sine expansion of the interpolation, from Eq.~\eqref{eq:fourier}.
}
\label{fig:single_tone_fit} 
\end{figure}

In the main text, we mention that we extract $E_\sigma$ from single-tone readout spectroscopy, which allows choosing proper spin-split energy in multi-ASQ experiments without two-tone spectroscopy.
We start from extracting $\it{f_\mathrm{r}}$ from Fig.~\ref{fig:spinT1} b), for both spin states, as shown in Fig.~\ref{fig:single_tone_fit} in red and orange dots.
With data for both spin states within $-0.5\Phi_0$ to $0\Phi_0$, we can deduce the entire readout curve between $-0.5\Phi_0$ to $0.5\Phi_0$ for both spin states owing to time-reversal symmetry. 
We apply a Fourier expansion to the discrete data points of form
\begin{equation}
    f(\varphi) = 0.5 A_0 +  \sum_{n=1}^N (A_n\cos{n\varphi} + B_n \sigma_z \sin{n\varphi} ) ~, \label{eq:fourier}
\end{equation}
with $0.5A_0=f_\text{r}$, period between $-\pi$ to $\pi$, and $A_1,B_1$ corresponds to $E_\mathrm{0},E_{\sigma}$ with pre-factors:
\begin{equation}
    \begin{split}
        E(\varphi) &= E_0\cos{\varphi} + E_\sigma \sigma_z \sin{\varphi} ~, \label{eq:full_odd_EPR}\\
        \frac{\partial^2E}{\partial \varphi^2} &= \delta f(\varphi)S^{-1}\left(\frac{\Phi_0}{2\pi}\right)^{2} =  -E_0\cos{\varphi} - E_\sigma \sigma_z \sin{\varphi}~.
    \end{split}
\end{equation}

The 0th and 1st order contribution extracted using Eq.~\ref{eq:fourier} are plotted in the cyan curve in Fig.~\ref{fig:single_tone_fit}.
We noticed that the two spin states remain degenerate until $0.05\Phi_0$ away from the Kramer's point, possibly due to sizable dephasing noise from hyperfine interaction.

On this bias point, $E_{\sigma}/h$ is extracted to be $\SI{255}{MHz}$ while the value from two-tone spectroscopy is $\SI{230}{MHz}$.
The spin-independent contribution to the dispersion is $E_\mathrm{0}/h = \SI{-291}{MHz}$.

\begin{figure*}
\centering
\includegraphics[width = \textwidth]{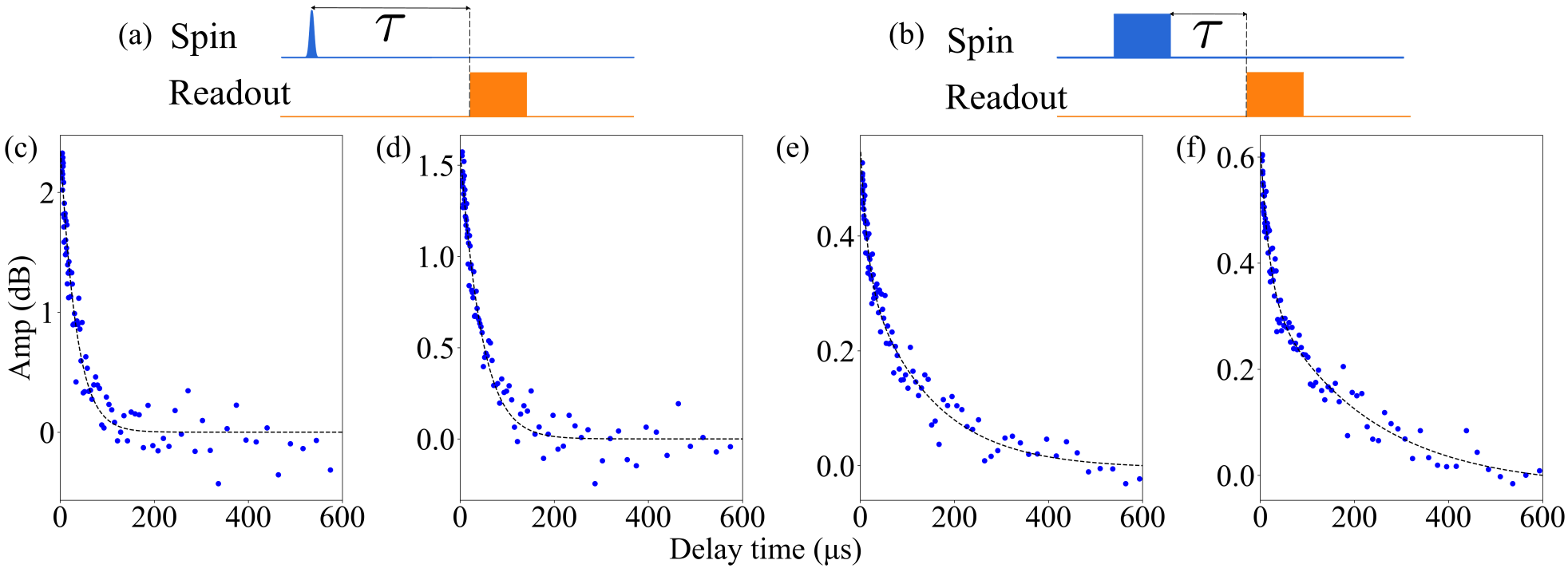}
\caption{ 
\textbf{Relaxation curve using short and long pulse.}
(a) Scheme of relaxation time measurement using a short pulse FWHM $=\SI{74}{\nano\second}$.
(b) Scheme of relaxation time measurement using a $=\SI{2}{\micro\second}$ long saturation pulse.
(c)-(d) Relaxation after driving the odd-parity state with short Gaussian pulse exhibits single-exponential decay, measured in the second cooldown at $V_\mathrm{g} = \SI{-1.786}{\volt}$. 
(c) At $-0.37\Phi_0$, relaxation time is $30.8\pm \SI{1.8}{\micro\second}$.
(d) At $-0.27\Phi_0$, relaxation time is $40.3\pm\SI{ 1.8}{\micro\second}$.
Note that in Fig.~\ref{fig:spinT1} (c), relaxation time is averaged over a few repeated scans.
(e)-(f) Relaxation after driving the odd-parity state with a long saturation pulse exhibits double-exponential decay behavior, measured in the first cooldown at $V_\mathrm{g} = \SI{-1.715}{\volt}$.
(e) At $0.21\Phi_0$, short decay time is $14.5\pm\SI{ 3.4}{\micro\second}$, long decay time is $138.2\pm \SI{21.6}{\micro\second}$.
(f) At $0.24\Phi_0$, short decay time is $18.2\pm \SI{3.3}{\micro\second}$, long decay time is $212.0\pm \SI{46.5}{\micro\second}$.
}
\label{fig:T1_ind} 
\end{figure*}

\subsection{Two-tone spectroscopy}\label{app:TT_spin}

\begin{figure}
\centering
\includegraphics[width = 1\columnwidth]{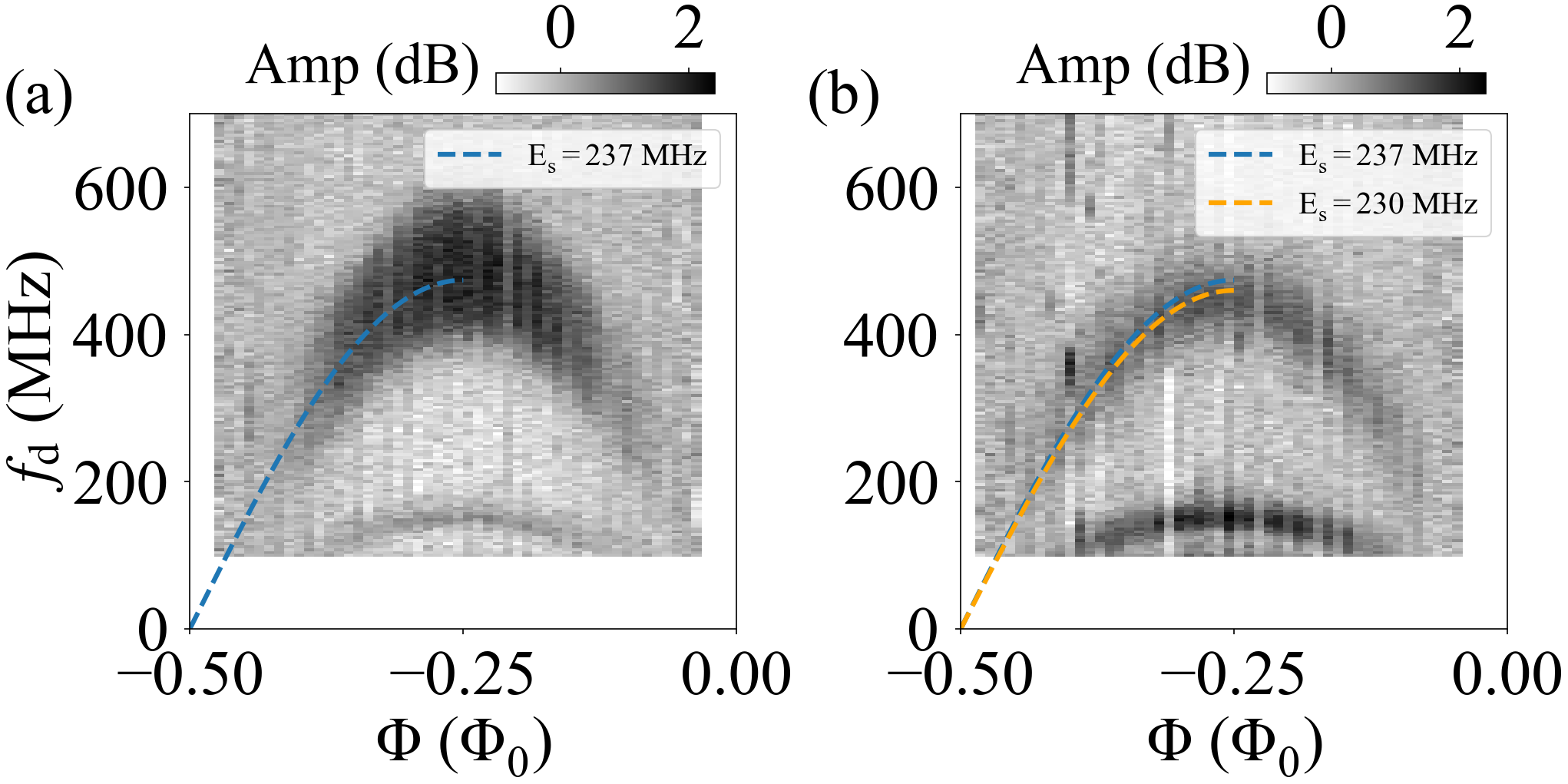}
\caption{ 
\textbf{Power broadening of $E_\sigma$.} (a) Two-tone spectroscopy taken together with Fig.~\ref{fig:spinT1} (c), with extracted $E_\sigma/h = \SI{237}{\mega\hertz}$. 
(b) Two-tone spectroscopy taken later with $\SI{18}{\deci\bel}$ lower drive power than in (a) for reducing high-power widened transition linewidth. 
}
\label{fig:SI_TT_power} 
\end{figure}

The two-tone spectroscopy in this work is collected using pulsed LO of $\SI{8}{\micro\second}$ duration time, defined by a homemade fast switch.
In Fig.~\ref{fig:spinT1} (b) we use low LO power and obtained $E_{\sigma}/h = \SI{230}{\mega\hertz}$. 
Fig.~\ref{fig:SI_TT_power} shows the difference between high and low power two-tone spectroscopy, with orange curve representing a faithful fit using $E_{\sigma}/h = \SI{230}{\mega\hertz}$ while blue curve representing $E_{\sigma}/h = \SI{237}{\mega\hertz}$. In Fig.~\ref{fig:TT_higher}, we use  $E_{\sigma}/h = \SI{230}{\mega\hertz}$.

In  Fig.~\ref{fig:SI_TT_power} (a) LO power is set at $\SI{-6}{\deci\bel}\text{m}$.
In  Fig.~\ref{fig:SI_TT_power} (b) LO power is set at $\SI{-24}{\deci\bel}\text{m}$.
Fig.~\ref{fig:spinT1} (c) is collected under high power short pulse. 
Thus we use $E_{\sigma}/h = \SI{237}{\mega\hertz}$ for relaxation time measurements.

\section{Relaxation time measurement} \label{app:T1meas}

Spin lifetime data presented in the main text is measured via pulsed drive, as shown in Fig.~\ref{fig:T1_ind} (a). 
The Gaussian pulse has FWHM of \SI{74}{\nano\second} with a total duration of \SI{160}{\nano\second} ($5\sigma$).
In the first cooldown, we collected relaxation time data via a $\SI{8}{\micro\second}$ flat top pulse drive. 
This results in a decay curve demonstrated in Fig.~\ref{fig:T1_ind} (e)-(f) that appears to exhibit a double-exponential decay.
The short timescale of the double-exponential decay is roughly consistent with the timescale measured in the pulsed experiment, though perhaps influenced by the saturation pulse.
Our conjecture for the source of the second, longer timescale is that the saturation pulse results in elevated temperatures of certain objects in the circuit (e.g. attenuators~\cite{maiti_cryogenic_2020}) or perhaps in the nanowire itself (e.g.,~\cite{ibabe_heat_2024}).
The double-exponential decay event appears much less frequently with the short Gaussian pulse.
We fit data in the first $\SI{250}{\micro\second}$ with a single-exponential function in Fig.~\ref{fig:spinT1} (c).

\section{Extended higher frequency transition two-tone spectroscopy} \label{app:higher_fre_TT}

\begin{figure}[b]
\centering
\includegraphics[width = 1\columnwidth]{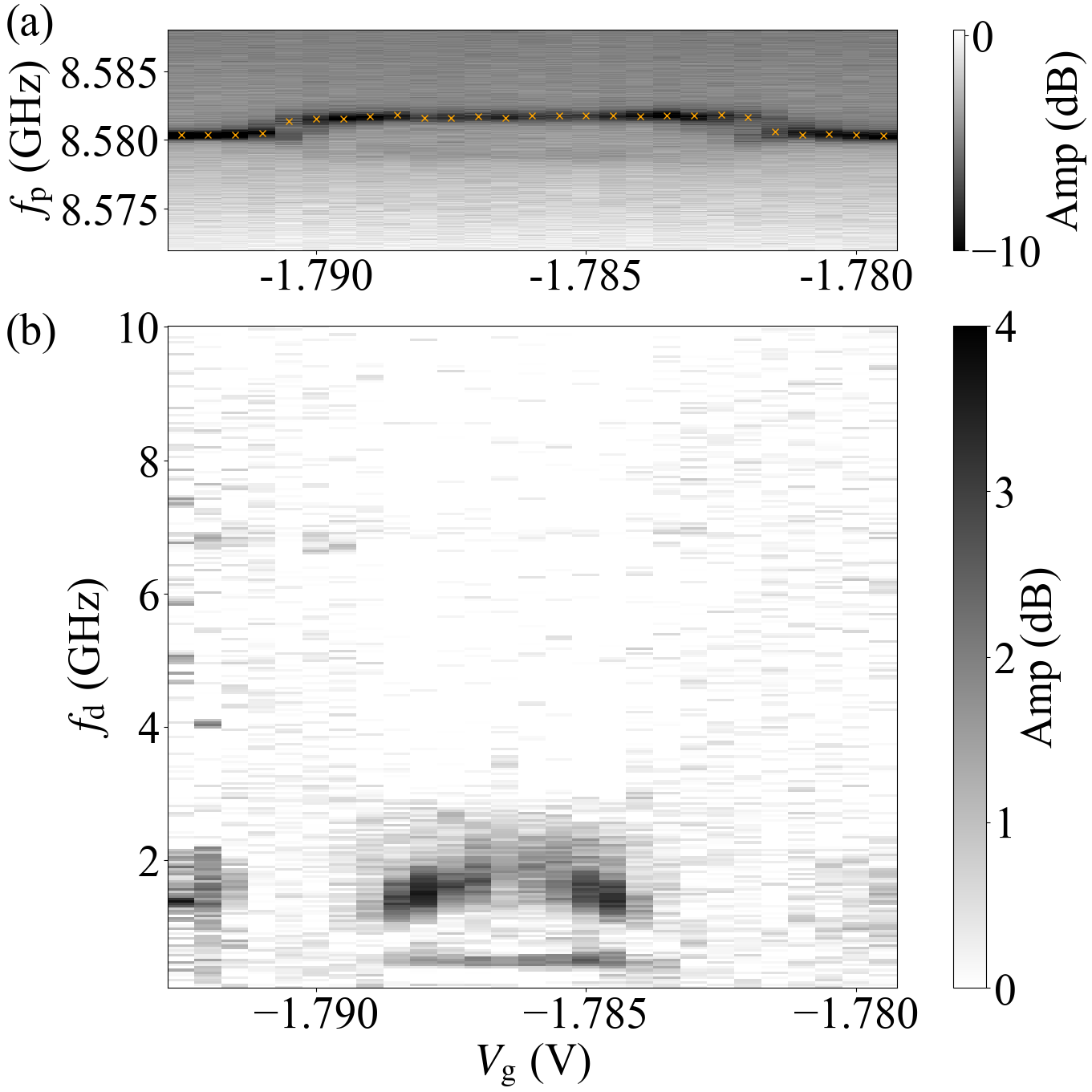}
\caption{ 
\textbf{Extended higher frequency transition two-tone spectroscopy.}
(a) Single-tone readout spectroscopy gate dependence at $\Phi = -0.25\Phi_0$. 
Orange dots show the readout frequency for two-tone spectroscopy scan.
(b) Two-tone spectroscopy gate dependence scan at $\Phi = -0.25\Phi_0$. 
Beyond the two transition curves shown in Fig.~\ref{fig:TT_higher} (e), no other transition is observed below $\SI{10}{\giga\hertz}$
}
\label{fig:SI_TT_weird_higher} 
\end{figure}

In Sec.~\ref{chp:Higher_ft}, we demonstrated a minimal model involving parity flipping by transitioning QP between the weak link and ancillary dot-like levels.
Here, we show two-tone spectroscopy up to $\SI{10}{\giga\hertz}$ in Fig.~\ref{fig:SI_TT_weird_higher} where only spin-flip transition and the parity flipping transition are captured.

\section{Data from additional bias points} \label{app:extra_bias}

\begin{figure*}[b]
\centering
\includegraphics[width = 1\textwidth]{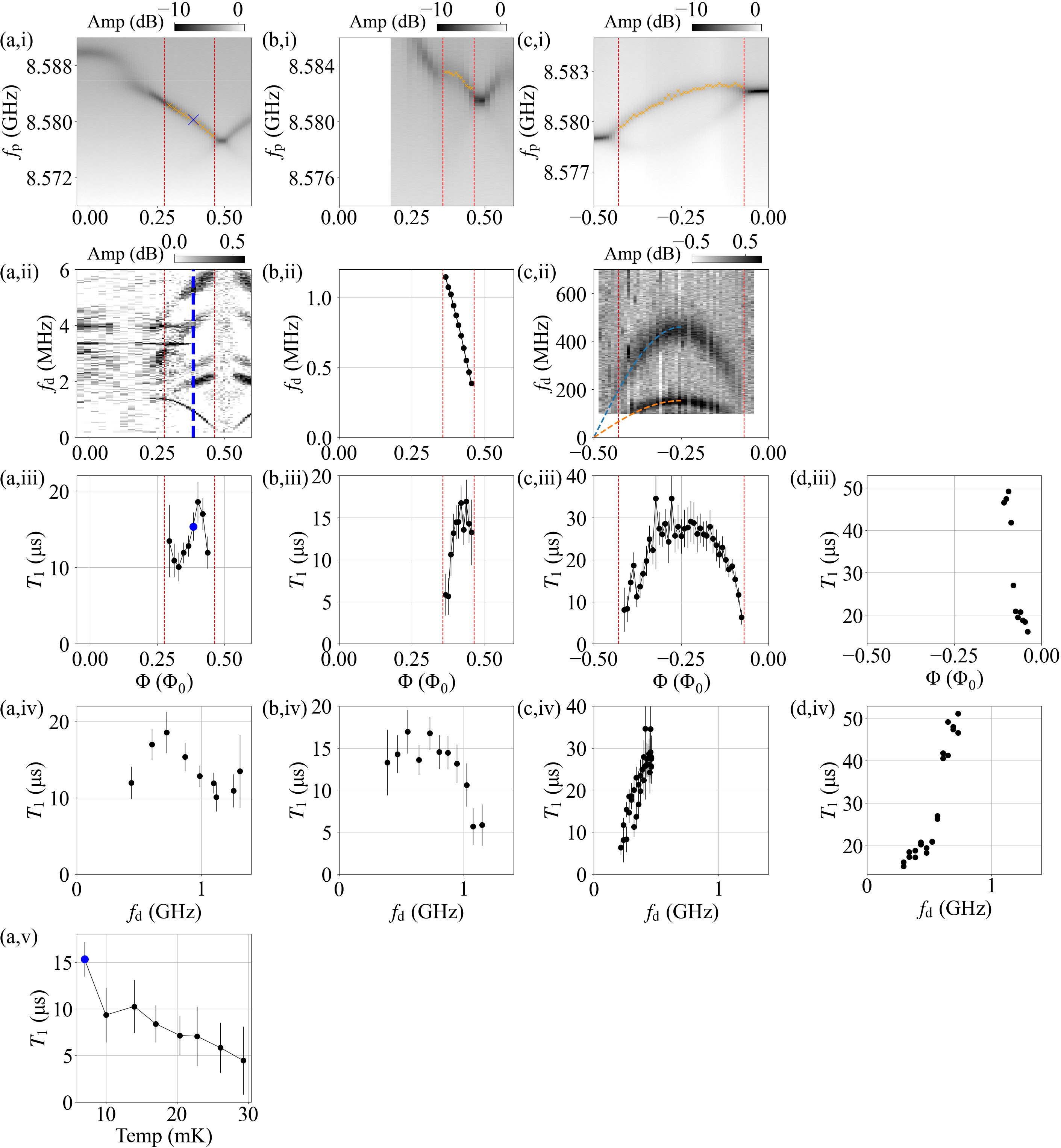}
\caption{ 
\textbf{Relaxation time, single- and two-tone spectroscopy scan from additional bias points.}
Column a: Data collected at $V_\mathrm{g}=\SI{-0.530}{\volt}$, third cooldown.
Column b: Data collected at $V_\mathrm{g}=\SI{-0.519}{\volt}$, third cooldown.
Column c: Data collected at $V_\mathrm{g}=\SI{-1.786}{\volt}$, second cooldown.
Column d: Data from Yale team's work\cite{hays_continuous_2020}.
Row \romannum{1}: Single-tone readout spectroscopy flux dependence, relaxation time data is collected at flux within red bars.
Row \romannum{2}: Two-tone spectroscopy and spin-flip transition frequency.
Row \romannum{3}: Spin relaxation time flux dependence.
Row \romannum{1}-\romannum{3}~ share the same x-axis.
Row \romannum{4}: Relaxation time spin-flip frequency dependence.
(a.\romannum{2}): Two-tone spectroscopy showing spin-flip transition with $E_\sigma = \SI{1.4}{\giga\hertz}$ and some higher frequency transitions. 
Even-odd parity switch takes place near $0.1-0.2 \Phi_0$.
(b.\romannum{2}): Two-tone spectroscopy showing spin-flip transition frequency with $E_\sigma = \SI{1.5}{\giga\hertz}$.
Even-odd parity switch takes place near $0.3-0.4 \Phi_0$.
(c.\romannum{2}): Two-tone spectroscopy with with $E_\sigma = \SI{230}{\mega\hertz}$.
(a.\romannum{5}): Temperature dependence of column a at $\Phi = 0.38\Phi_0$. 
}
\label{fig:SI_T1_fs_whays} 
\end{figure*}

In this section, we presented relaxation time flux dependence from a few bias points, shown in Fig.~\ref{fig:SI_T1_fs_whays}.
Our convention with the panel labels is that columns (a,b,c,d) refer to different bias points while row labels (i,ii,iii,iv,v) refer to different measurements. 
Specifically for the columns
\begin{enumerate}[label=\alph*]
    \item Third cooldown, $V_\mathrm{g} = \SI{-0.53}{\volt}$
    \item Third cooldown, $V_\mathrm{g} = \SI{-0.519}{\volt}$
    \item Second cooldown, $V_\mathrm{g} = \SI{-1.786}{\volt}$ (same as in Fig.~\ref{fig:spinT1} and Fig.~\ref{fig:TT_higher} of main text)
    \item Data from Hays, et al.~\cite{hays_continuous_2020}
\end{enumerate}
and the rows
\begin{enumerate}[label=\roman*]
    \item Single-tone readout spectroscopy, in which the orange crosses show the flux on which we collected relaxation time data
    \item Two-tone spectroscopy and spin-flip transition frequency
    \item Relaxation time data vs magnetic flux
    \item Relaxation time data vs spin transition frequency 
    \item Temperature-dependence of relaxation time, for column (a) only at $\Phi = 0.38\Phi_0$
\end{enumerate}
To reduce text clutter, we will drop the "Fig." label when referring to panels from Fig.~\ref{fig:SI_T1_fs_whays} in this section only.

In (a.\romannum{2}), we recorded another two-tone spectroscopy with high-frequency transitions.
The transitions are observed only in the flux regions where the ground state is in odd-parity.
It is unclear to us whether these are transitions to higher odd-parity levels or involve parity flipping similar to the mechanism explained in Sec.~\ref{chp:Higher_ft}.

In (a.\romannum{3}) and (b.\romannum{3}), maximum relaxation time are lower than (c.\romannum{3}) and (d.\romannum{3}), which is possibly due to even-odd parity switch in adjacent flux, as demonstrated in Fig.~\ref{fig:spinT1} (c) where relaxation time drops when closer to even-odd parity switch.
Relaxation time is also not saturated around $-0.25\Phi_0$ as in (c.\romannum{3}) due to a similar reason that both spin relaxation rate and QP transition relaxation rate contribute to relaxation rate.
(c.\romannum{3}) shows relaxation time flux dependence line cut at $V_\mathrm{g} = \SI{-1.786}{\volt}$.
The relaxation time is at a maximum near $-0.25\Phi_0$.
The decreasing trend towards $0\Phi_0$ is similar to what has been observed in (d.\romannum{3})~\cite{hays_continuous_2020}.

To better understand the origin of the relaxation, we plot the relaxation time versus the transition frequency.
(c.\romannum{4}) is taken at gate voltage bias $V_\mathrm{g}=\SI{-1.786}{\volt}$ where relaxation rate is minimum among the whole gate range in Fig.~\ref{fig:spinT1}(c).
(c.\romannum{4}) and (d.\romannum{4}) has relaxation time $\SI{30}{\micro\second}$ at $\SI{470}{\mega\hertz}$/$\SI{580}{\mega\hertz}$ respectively.
We thus conjecture that the spin lifetime of these weak links hosted on InAs nanowire with Al shell may be confined by a similar frequency-dependent loss mechanism, hyperfine interaction, for instance.

Fig.~(a.\romannum{5}) shows the temperature dependence of the relaxation time on flux bias marked blue in column a.
In \cite{hays_continuous_2020} (column d), the spin lifetime does not have strong temperature dependence up to \SI{100}{\milli\kelvin}.
It is unclear what contributes to the difference between the two experiments.

\section{Estimation on QP poisoning}~\label{app:T_poisoning}

In our single-tone spectroscopy experiments, we did not observe opposite-parity states in the same regions of the parameter space (outside of clear transition regions).
In general, this is a qualitative indication that QP poisoning has been mitigated to some extent, but the lack of a single-shot readout makes a direct assessment of the poisoning rate challenging. 
Here, we give a coarse estimate for the upper bound of the population  $P$ of the state with opposite parity to the ground state by estimating the smallest resolvable peak in single-tone spectroscopy.

\begin{figure}[b]
\centering
\includegraphics[width = 1\columnwidth]{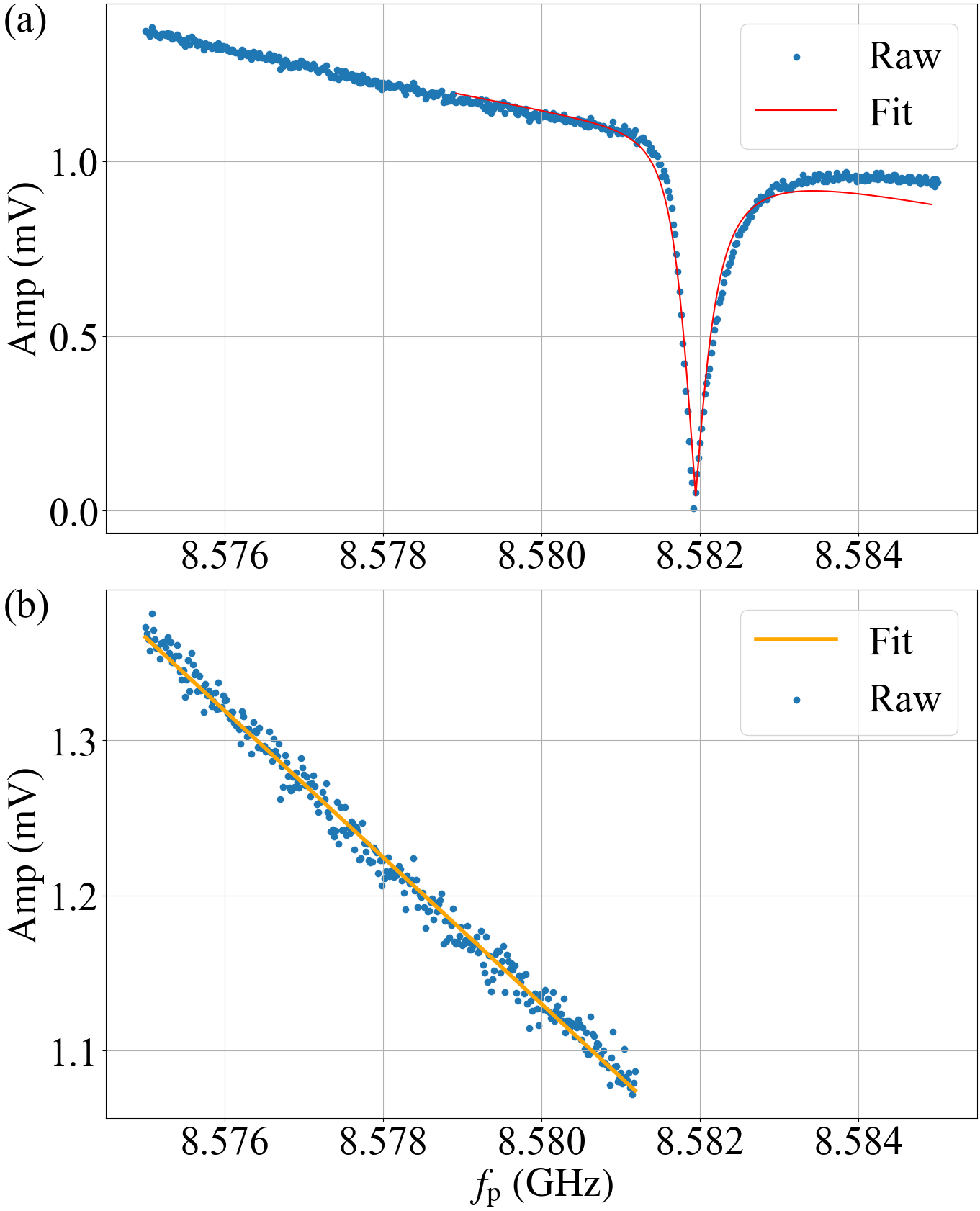}
\caption{ 
\textbf{Resonance peak fit and signal standard deviation extraction.}
(a) Fitting of the resonance peak at $-0.15\Phi_0$, $V_\mathrm{g} = \SI{-1.775}{\volt}$, even manifold, second cooldown.
The resonance peak is fitted after a linear background subtraction using DCM method~\cite{khalil_DCM_2012}.
Fitted parameter: $\kappa_c/2\pi = \SI{0.323}{\mega\hertz}$, $\kappa_i/2\pi = \SI{0.317}{\mega\hertz}$ and $f_\text{r} = \SI{8.582}{\giga\hertz}$.
(b) Linear fit and signal STD extraction of background signal between $\SI{8.575}{\giga\hertz}$ to $\SI{8.5812}{\giga\hertz}$.
We assume a linear frequency-dependent background in this small scanning range and calculate $s$ accordingly.
}
\label{fig:SI_Resonator_fit} 
\end{figure}

First, we characterized the expected resonator lineshape by a fit to a standard scattering model in Fig.~\ref{fig:SI_Resonator_fit}(a) (data from Fig.~\ref{fig:SI_even_singletone} at $-0.15\Phi_0$). 
From this we obtain a coupling loss $\kappa_c/2\pi = \SI{0.323}{\mega\hertz}$ and an internal loss $\kappa_i/2\pi = \SI{0.317}{\mega\hertz}$.

Then, we characterized the noise level of the measurement of the reflection amplitude $A$.
Taking a frequency range outside the peak area, we assume a linear frequency-dependent background $A_{BG} = A_0 + \alpha f_p$, where $A_0$ is the frequency-independent part, $f_p$ is the measurement frequency, and $\alpha$ is the slope with respect to that frequency.
We then obtain a standard deviation
\begin{equation}
    s = \sqrt{\frac{1}{N}\sum_{n=1}^N \left(A - A_\text{BG}\right)^2}~.
\end{equation}
We obtain standard deviation $s=\SI{8.32}{\micro\volt}$ with mean signal amplitude of $ \langle A_{BG}\rangle  = \SI{1.22}{\milli\volt}$ in the frequency range of interest.    

Assuming we can distinguish a peak amplitude of $n$ standard deviations, the upper bound for $P$ can be calculated by solving:
\begin{equation}
    \frac{\kappa_i-\kappa_c}{\kappa_i+\kappa_c}P +(1-P) = \frac{\langle A_{BG}\rangle-n s}{\langle A_{BG}\rangle}~.
\end{equation}
Then, taking a conservative value of $n=5$ for a strongly measurable signal, we obtain the estimate $P < 0.0068$.
With our measurement time of \SI{5}{\micro\second}, we therefore bound the poisoning rate to be longer than \SI{0.7}{\milli\second}.

\section{Thermal cycle and data collection}

Data presented in this work are from three cooldowns. 
Fig.~\ref{fig:singletone}, Fig.~\ref{fig:T1_ind} (e)-(f) were collected during cooldown 1.
Fig.~\ref{fig:spinT1}, Fig.~\ref{fig:TT_higher}, Fig.~\ref{fig:SI_even_singletone}, Fig.~\ref{fig:single_tone_fit}, Fig.~\ref{fig:T1_ind} (c)-(d), Fig.~\ref{fig:SI_TT_power}, Fig.~\ref{fig:SI_TT_weird_higher}, and Fig.~\ref{fig:SI_T1_fs_whays} column (c)  were collected during thermal cycle 2.
 Fig.~\ref{fig:SI_T1_fs_whays} column (a)-(b)  were collected during thermal cycle 3.

\balancecolsandclearpage

\bibliography{vf-references,hl-references}
\end{document}